\documentclass[twocolumn,superscriptaddress,pra,showkeys,nofootinbib,floatfix,raggedbottom]{revtex4-2}

\usepackage{amsmath, amssymb, amsthm}
\usepackage{mathtools}
\usepackage{xurl}
\usepackage{graphicx}
\usepackage{booktabs}
\usepackage{hyperref}
\hypersetup{
    colorlinks=true,
    linkcolor=blue,
    citecolor=blue,
    urlcolor=blue
}

\newtheorem{theorem}{Theorem}
\newtheorem{lemma}{Lemma}
\newtheorem{proposition}{Proposition}

\allowdisplaybreaks

\makeatletter
\newcommand{\tabnotes}[1]{%
  \par\vspace{0.5em}%
  \parbox{\columnwidth}{\footnotesize
    \leftskip\z@skip \rightskip\z@skip \parfillskip\@flushglue \parindent\z@
    \textit{Notes:} #1}}
\makeatother

\newcommand{\runhead}[1]{\par\addvspace{\smallskipamount}\noindent\textbf{#1.}}

\hypersetup{
    colorlinks=true,
    linkcolor=blue,
    citecolor=blue,
    urlcolor=blue
}

\begin{document}

\title{How Much Screening Buys a Second? Speed, Selection, and the Maker's Spread}

\author{Aleksei Adadurov}
\affiliation{nuconstruct \ \textnormal{\texttt{team@nuconstruct.xyz}}}

\author{Eduard Airapetian}
\affiliation{University of Wisconsin--Madison}

\author{Anton Chtepine}
\author{Antero Eloranta}
\author{Boris Pletenev}
\author{Andrei Sebyakin}
\author{Arsenii Valitov}
\affiliation{nuconstruct \ \textnormal{\texttt{team@nuconstruct.xyz}}}

\date{\today}

\begin{abstract}
We study the spread a liquidity provider needs to break even against counterparties that can exploit a stale quote, and we place two defenses on a single scale: repricing sooner and screening informed flow. A break-even condition pins down the exact screening rate at which a slower maker matches a faster unscreened venue: the selection that buys back the extra staleness. The result requires only an integrable price innovation and therefore holds under jumps, atoms, and infinite variance. Predictable drift is a centering problem: it widens a spot-centered quote, but its cost vanishes once the quote is recentered on the expected reference. In four USDC-WETH pools benchmarked against Binance ETH-USDT from June to August 2026, the one-second benchmark innovation is sharply non-Gaussian: 73\% of seconds show no price change. Cutting cadence from twelve seconds to one sheds 84.5\% of the expected shock, so a twelve-second maker must screen at least 84.5\% of informed flow to match a one-second venue, far above the 71.1\% that a Gaussian model implies. In trending markets, the stale pool is picked off at execution: with-trend fills obtain an effective spread 11.1\,bps lower than against-trend fills, with the same sign in all four pools.
\end{abstract}

\keywords{automated market makers, adverse selection, market microstructure, order-flow toxicity}

\maketitle

\section{Introduction}\label{sec:introduction}

A liquidity provider (LP) who posts a firm two-sided quote has written an option on the asset. Traders who know where the price is heading exercise it against the maker, while everyone else trades by chance. The maker recovers what the informed take from the spread it charges the rest. The competitive spread is the width at which the two balance: losses to better-informed counterparties are repaid by earnings from everyone else.

Automated market makers, or AMMs, make this option unusually stark. A constant-function pool quotes a deterministic price off its reserves and refreshes only when someone trades against it. Between trades, the quote is stale, and on a chain that settles in discrete blocks, it can stay stale for the whole block, roughly twelve seconds on the network we study. With no last look and no way to pull a posted quote, an arbitrageur watching a faster venue picks the pool off the moment the external price clears fees. That is, the pool pays, block after block, for a price the world has already left behind.

A maker has two defenses: shorten the time its quote is exposed, or screen the flow and decline the orders most likely to be informed. Which defense is available depends on the venue. A permissionless pool must fill every order that clears its invariant, so its only lever is exposure; a proprietary or curated maker can do both. Our question is how much a second of reduced staleness is worth against a unit of screening, and when one substitutes for the other.

We answer this question with a model of the competitive half-spread built from three ingredients: the price innovation over the maker's exposure horizon, the chance a counterparty is informed, and the fraction of informed flow screened out. A single object, the expected excess of the innovation over the spread, prices both defenses. When a fast reference layer is compared with a slower pool, there is a critical screening rate at which the slow pool, by screening alone, matches the fast layer's spread; this is the selection that buys back the extra staleness.

Three features organize the paper. First, the threshold is scale-free: under a driftless diffusion it depends only on the ratio of the two horizons and the informed share $\theta$, not on volatility, fee, or trade size. Second, the threshold's response to toxicity is governed by the expected loss on an executed informed order, so a fat-tailed innovation and a Gaussian one of equal variance imply very different spreads and screening thresholds: the shape of the tail, not just its width, is what the maker prices. Third, drift is a centering problem: a predictable price component tilts which side gets picked off and widens a quote held at the current reference, but a maker that centers its quote on the expected reference recovers the driftless spread, as Section \ref{sec:model-drift} makes precise.

We then take these predictions to a panel of USDC-WETH pools benchmarked against a faster centralized venue. The benchmark innovation is sharply non-Gaussian at both the one-second and the block horizon, and we use it to compute the screening threshold, to show how sensitive that threshold is to the shape of the tail, and to test the model's directional prediction that a stale quote is picked off on the side toward which the reference is moving. The magnitudes are reported in Section \ref{sec:analysis}.

The rest of the paper proceeds as follows: Section \ref{sec:literature} places the model in the existing literature; Section \ref{sec:model} develops the model of the spread, the screening threshold, and their comparative statics; Section \ref{sec:model-drift} adds predictable drift; Section \ref{sec:analysis} presents the empirical calibration and tests; and Section \ref{sec:conclusion} concludes.

\section{Related Literature}\label{sec:literature}

This paper draws on three literatures that have developed largely separately: the microstructure theory of adverse selection and the spread, the market-design literature on speed and batch auctions, and work on automated market makers, loss-versus-rebalancing, and extractable value. The model combines the break-even logic from the first, the focus on quote exposure from the second, and the empirical object, a stale on-chain quote adversely selected against a centralized reference, from the third.

\subsection{Adverse selection and the break-even spread}

The idea that the spread compensates a market maker for losses to better-informed counterparties is due to \citet{bagehot1971}. \citet{copelandgalai1983} turn it into a pricing problem in which posting a two-sided quote is writing a straddle to the informed, and an informed trader whose edge does not exceed the spread does not trade. That participation margin is the truncation at the center of our model, which we close with the zero-expected-profit condition of \citet{glostenmilgrom1985}, who show that a positive spread survives the removal of every inventory and order-processing cost. \citet{stoll1978} and \citet{hostoll1981} give the competing inventory account, in which a dealer's position shifts the midpoint of the quote rather than its width, and resetting every quote to the contemporaneous reference removes that channel by construction. The same distinction between the center and the width of a quote organizes our treatment of predictable drift. A known drift is a location shift that a maker free to choose its center can absorb without widening the spread, and it widens the spread only when the center is held at the current reference.

\citet{kyle1985} fixes the benchmark for the price process. In our benchmark, the reference price shares his martingale property, and the model departs from his setting in two ways that matter for what can be proved. The maker cannot revise inside an execution interval, so the relevant risk is the accumulated innovation over that interval rather than the response to a single batch. And we require only that the innovation be integrable, so jumps, atoms, and infinite variance are all admitted, with a predictable component entering as a known location shift rather than as serial correlation in the innovations. What is new relative to this line is the object being priced. Here the innovation is the increment of the reference price over one execution interval, so its dispersion is set by the repricing cadence; a venue that reprices every second can be compared with one that reprices every block on a single scale, and the screening that equalizes their spreads becomes a well-defined quantity.

\subsection{Measuring flow toxicity}

The empirical counterpart of toxicity is the probability that an order is informed. \citet{ekop1996} recover it by maximum likelihood from daily buy and sell counts (PIN), and \citet{elo2012} adapt the idea to high-frequency data as an order-imbalance statistic over volume buckets (VPIN). \citet{andersenbondarenko2014} find VPIN's forecasting power to be largely a mechanical consequence of its correlation with trading intensity, a reading that \citet{elo2014rejoinder} dispute. The netting measure we construct below belongs to this family but is benchmarked against a randomized null and treated as an upper bound, because parent-order splitting makes even uninformed flow arrive one-sided within a short batch. A related concern rules out the obvious alternative: against a martingale reference, a trade's markout is as likely to be favorable as unfavorable, whatever the informed share. We therefore measure the stale-quote loss in the execution price, through the effective spread of \citet{bessembinderkaufman1997}, and contrast it with the permanent price impact of \citet{hasbrouck1991} evaluated at a fixed horizon after the fill.

\subsection{Cadence, latency, and batch auctions}

The discrete execution interval at the center of our model has the timing of a frequent batch auction in the sense of \citet{budishcramtonshim2015}, run at subslot cadence, that is, at one-second intervals within Ethereum's twelve-second slot. \citet{aquilinabudishoneill2022} measure the rents that continuous trading creates: on FTSE\,100 stocks, latency-arbitrage races account for roughly 20\% of volume and impose a tax of about half a basis point, a third of the effective spread. That figure is the traditional-market counterpart of the staleness cost this paper measures for AMMs. \citet{budishleeshim2024} explain why incumbents do not adopt the fix, since exchanges earn rents from selling speed. \citet{wahwellman2013} and \citet{baldaufmollner2020} reach the same design conclusion as \citet{budishcramtonshim2015}, and Wah and Wellman add that a call market yields efficiency gains from aggregating orders over and above those from neutralizing latency arbitrage. That is why the share of flow that offsets within an interval is worth measuring even though our pricing model, which prices each order at the margin, does not net orders against one another.

On the choice of cadence, \citet{menkveldzoican2017} show that a faster exchange lets a market maker refresh stale quotes sooner but also strips away the protection of queued liquidity orders, so it widens the spread only when news arrival is high relative to liquidity-trader arrival. Our cadence comparison isolates the first channel by holding the composition of potential flow fixed across venues, and their result is why the second channel has to be treated as a separate effect rather than folded into the same comparison. \citet{frickegerig2018} find that market quality is maximized at an interior clearing interval of a few seconds for a typical S\&P\,500 stock, the closest quantitative analogue to a one-second subslot. \citet{brolleycimon2020} model a latency delay as a device that concentrates informed trading on the fast venue, an assignment our setting inverts. On-chain, \citet{milionis2023fees} show that faster blocks mechanically reduce provider losses because arbitrage profit scales with the fraction of blocks that present an opportunity, while \citet{adadurov2026subslots} find, in calibrated simulations, that one-second subslots raise the count of arbitrages between centralized and decentralized exchanges (CEX-DEX arbitrages) by 535\%. Together, these separate the two effects of speed: shorter exposure per interval, which the model prices at a fixed composition of potential flow, and a venue that is worth attacking more often, which changes that composition and is left to measurement.

\subsection{Segmentation and screening}

Our comparison turns on one asymmetry. A proprietary maker can decline the orders it judges likely to be informed, and a shared venue that must accept whatever arrives cannot. The literature supplies both the mechanism and the evidence that the asymmetry is real. \citet{battalioholden2001} show that externally verifiable order characteristics let a wholesaler keep the retail flow it can identify cheaply and route the rest onward, so the venue of last resort faces a more informed residual. This is cream-skimming, and it is a stable equilibrium rather than a transitional one. Their distinction between characteristics a third party can verify and those only the broker can is the distinction between a screen's false positives and its false negatives, and \citet{bjzz2021} show that a residual pricing feature, sub-penny execution, identifies retail flow in public data and can be validated against an audit trail. \citet{zhu2014} obtains the same sorting without anyone rejecting an order, since informed traders crowd one side of a midpoint dark pool and migrate to the exchange when they fail to execute. Together, they make the point our screening parameter encodes: a signal about who is informed changes a maker's losses only if the maker acts on it, and observing it without acting changes nothing. \citet{degrysedejongvankervel2015} find that dark trading raises price impact on lit venues, \citet{madhavan1995} shows that without mandatory disclosure such segmentation persists, and on Ethereum \citet{oz2024builders} and \citet{wu2025cexdex} find that exclusive order flow is what makes builders and searchers profitable. The closest live counterpart to the screened maker we model is the professional market maker quoting from private inventory behind a solver auction, whose delivered execution quality \citet{yuminaga2025solver} find varies materially with the composition of the solver set. Because those makers and the pools they compete with reprice at the same block cadence, the price improvement they deliver is a spread premium at equal speed, the object our equal-cadence comparison isolates.

A different route to protecting makers is to show them part of the incoming flow before they quote rather than to let them reject it, and we do not model that channel. The evidence on it cuts both ways. \citet{paganoroell1996} show that greater transparency lowers the costs of the uninformed, whereas \citet{madhavanporterweaver2005} find that trading costs rose when the Toronto Stock Exchange began displaying its order book, because a displayed order is a free option to whoever sees it. \citet{foucaultmoinastheissen2007} and \citet{rindi2008} show that anonymity can instead retain the informed traders who supply liquidity most cheaply, and the benign case, sunshine trading in \citet{admatipfleiderer1991}, works only because the disclosed orders are credibly uninformed.

\subsection{Automated market makers and loss-versus-rebalancing}

The unscreened benchmark at block cadence is the constant-function liquidity provider, whose quote can be re-anchored only by an on-chain transaction. \citet{angeris2021uniswap} show that arbitrage confines the pool price to a fee-determined band around the reference, which is the residual mispricing our resetting assumption abstracts from, and \citet{evansangerischitra2021} show that fees widen the no-trade band and so trade price accuracy against provider protection. \citet{milionis2022lvr} decompose provider returns into a rebalancing component and a residual equal to fee income minus an arbitrage loss, loss-versus-rebalancing, which is quadratic in volatility and depends on the pool only through the curvature of its bonding function. Our object is instead the first absolute moment of the interval innovation, truncated at the spread. It is defined for any integrable innovation rather than requiring a diffusion, and because we price marginal trades, it sets curvature aside, which is the margin \citet{capponijia2021} study when they trade the severity of the arbitrage problem against takers' surplus. \citet{leharparlour2023} document stable pools and no long-lived arbitrage against Binance, the reference this paper also adopts. \citet{cartea2024pl} derive an optimal concentrated-liquidity range and skew it on a predictable drift, the provision-side counterpart of recentering a quote on the expected reference. Two design responses are comparable to ours. \citet{adams2024amamm} auction the right to manage a pool, and \citet{mcmenamin2022diamond} auction the arbitrage opportunity itself. Both accept that staleness will be arbitraged and route the proceeds back to providers, whereas we ask how much of it faster repricing removes at source and what screening rate substitutes for the difference. \citet{milionis2023myerson} provide the benchmark for any such design, showing that the optimal incentive-compatible AMM has a jump in its demand curve, a bid-ask spread, from adverse selection when asymmetry is large and from monopoly pricing when it is small.

\subsection{CEX-DEX arbitrage}

The informed trader of our model, who observes the updated reference before execution, is the CEX-DEX searcher. \citet{daian2020} documented bots bidding up fees in priority gas auctions and named the phenomenon miner extractable value, now maximal extractable value, and \citet{qinzhougervais2022} measured \$540 million of it over 32 months, of which \$277 million came from DEX arbitrage. Our identification treats the centralized reference as predetermined with respect to any single pool's flow, an assumption that is supported by \citet{makarovschoar2020}, who show that a common component of signed volume explains 80\% of bitcoin returns across 34 exchanges, and that we verify directly with a slot-level lead-lag regression in the price-discovery tradition of \citet{hasbrouck1995}. Our use of a centralized midpoint as the benchmark follows \citet{leharparlour2023}, \citet{wu2025cexdex}, and \citet{adadurov2026subslots}.

The model holds uninformed volume fixed, so any response of benign flow to execution costs sits outside it and would require an elasticity that the literature cautions against recovering from a venue cross-section. \citet{foucaultmenkveld2008} find liquidity supply strongly elastic to fees, \citet{colliardfoucault2012} and \citet{malinovapark2015} show that only the total fee, not its maker-taker breakdown, affects fill rates and welfare, \citet{hasbrouckriverasaleh2023} show that in an AMM a higher fee can raise volume by attracting inventory, and \citet{barbonranaldo2024} find that gas makes centralized venues cheaper for all but large trades. Depth and fee tier jointly determine a pool's volume and its effective spread, so a cross-section of pools cannot by itself identify how takers respond to price.

\section{Theoretical Model}\label{sec:model}

We study the spread required to compensate a liquidity provider for trading against orders that can exploit an outdated quote. Two features of the trading arrangement determine this exposure: the time for which the quote remains binding as the external reference price changes, and the maker's ability to limit execution by informed counterparties. We represent these features by a quote-exposure horizon and a screening rate. Holding potential order flow fixed allows us to compare their separate effects and determine the screening rate that offsets a longer exposure.

The model combines the option interpretation of posted quotes in \citet{copelandgalai1983} with the competitive adverse-selection pricing logic of \citet{glostenmilgrom1985}. Its object is the marginal execution cost of small trades. Liquidity providers are risk-neutral, and inventory costs, funding costs, and the price impact of individual orders are excluded. Within this setting, we characterize the required break-even spread, the screening threshold that equalizes spreads across exposure horizons, and the remaining spread premium of an unscreened venue when horizons coincide.

\subsection{Reference prices and execution timing}

Time is measured in fixed base intervals. Let $P_t>0$ denote the external reference price and let $p_t=\log P_t$. At integer times, define the one-period log-price innovation by
\[ \varepsilon_t := p_{t+1}-p_t\]
We assume
\[\varepsilon_t \overset{\mathrm{i.i.d.}}{\sim} F,
    \quad
    \mathbb{E}[\varepsilon_t]=0,
    \quad
    0<\mathbb{E}[|\varepsilon_t|]<\infty\]

Future increments are independent of the public price information
$\mathcal{F}_t$ available at time $t$. Thus the log price is a
martingale relative to this filtration. This is the benchmark
without predictable drift; Section \ref{sec:model-drift} treats
nonzero drift separately.

The reference price is external to the modeled venue, so an individual order does not change its law. The zero-mean assumption rules out predictable changes in the log reference price over the base interval. Independence and stationarity provide a common innovation law from which exposures at different horizons are constructed. Requiring a finite first absolute moment ensures that the expected trading losses below are finite, while allowing discrete innovations and distributions without a finite variance. One admissible continuous-time specification of these assumptions is
\[
    dp_t=\sigma\,dW_t+d\widetilde{J}_t,
\]
where $\sigma\geq 0$, $W_t$ is a Brownian motion, and
$\widetilde{J}_t$ is an independent, integrable, mean-zero Lévy jump process, non-degenerate when $\sigma = 0$ (so that $\mathbb{E}|\varepsilon_t| > 0$). The results below require only the stated properties of the discrete increments.

At the start of an interval, the maker sets a quote using the current reference price. The quote remains binding until execution at the end of the interval. Immediately before execution, informed traders observe the updated reference and can submit an order when trading against the standing quote is profitable. The maker cannot incorporate that update before these orders execute. The information advantage therefore consists of an ability to act on a reference-price change while the quote remains fixed; it can arise even when the updated reference is publicly observable. This is the stale-quote mechanism emphasized by \citet{budishcramtonshim2015}.

We call $h$ the quote-exposure horizon: it is the elapsed time between the reference observation used to set the quote and the execution opportunity. A shorter horizon means more frequent effective repricing. Mapping $h$ to a particular venue requires identifying when its quote becomes binding and when an informed order can execute against it. A nominal block duration or update frequency represents this horizon only when it matches those events. Appendix \ref{app:execution-ages} allows opportunities to occur at different quote ages within an update interval.

After execution, the quote is reset to the contemporaneous reference. Exact resetting gives the venues a common starting point and isolates the discrepancy accumulated during the next exposure interval. This is a maintained benchmark assumption. With proportional trading fees, arbitrage in a conventional AMM can leave its marginal price inside a no-arbitrage band around the reference \citep{angeris2021uniswap}. Such residual discrepancies would become an additional state variable in a model of the pool's price dynamics.

For an interval beginning at $t$, the relevant innovation is
\[ Z_{t,h}:=p_{t+h}-p_t = \sum_{k=0}^{h-1}\varepsilon_{t+k}\]

By stationarity, we suppress the time index and write $Z_h$ for integer $h$,
\[ Z_h\sim F^{*h}, \quad \mathbb{E}[Z_h]=0, \]
where $F^{*h}$ denotes the $h$-fold convolution of $F$.

The reference execution layer is assumed to implement the quote-setting
and execution rules above at a fixed integer horizon $h_0\geq1$,
without screening. A slower unscreened benchmark has integer
horizon $h_C\geq h_0$. A proprietary maker, which may also screen
its flow, has integer horizon $h_P\geq h_0$. These horizons describe effective exposure
to an unchanged quote; their interpretation for particular market
designs is left to the empirical application.

\subsection{Quotes and the interpretation of the spread}

Let $s\geq 0$ denote the symmetric half-spread in log-price units.
A quote centered on $P_t$ has bid and ask prices
\[P_t^{\mathrm{bid}}=P_t e^{-s}, \quad P_t^{\mathrm{ask}}=P_t e^{s}\]

For an actively quoted venue, $s$ directly describes the quoted half-spread. For an AMM, we interpret it as a reduced-form marginal execution wedge. Under the usual convention that a proportional fee $f$ is deducted from the input amount, the fee-induced marginal log half-spread is
\[s=-\log(1-f)\simeq f\]

The spread parameter puts actively posted quotes and the marginal execution wedge of an AMM in common units. For an input fee $f$, the marginal bid and ask multipliers are $1-f$ and $1/(1-f)$, which yield the stated log half-spread \citep{angeris2021uniswap}. This correspondence concerns infinitesimal trades at a given marginal pool price. It does not determine the amount an arbitrageur would trade along the pool's curve or the execution cost of a finite order.

Let $D_i\in\{-1,+1\}$ denote an order's direction, with $D_i=+1$
for a customer buy and $D_i=-1$ for a customer sell.
We measure trading profits per unit of initial notional and
use the first-order approximation
\[\Pi_i=s-D_i Z_h\]

The first-order payoff makes the loss on an informed order equal to its directional log-price advantage in excess of the half-spread. Every spread comparison below uses this payoff convention. Its quantitative interpretation requires price changes and spreads to be small enough for the approximation to be accurate. Integrability of log-price innovations establishes the existence of expected losses in this model; it does not by itself control approximation errors for exact payoffs expressed in price levels.

\subsection{Potential order flow and participation}

Potential flow arrives at a constant rate, independently of price innovations. Normalize expected potential order notional per unit of time to one. An execution interval of length $h$ therefore receives expected potential notional $h$.

Orders are equally sized in initial-notional terms.
A fraction $\theta\in(0,1)$ of potential flow is informed, and the remaining fraction $1-\theta$ is uninformed. We write $I$ and $U$ for the events that a given order is informed or uninformed, respectively. Thus, $\theta$ describes potential flow before screening
and before traders decide whether to execute. The informed share of executed orders is an outcome of both screening and the spread. We hold the potential arrival rate and $\theta$ fixed across venues to isolate the effects of the trading arrangement.

Uninformed traders execute for liquidity reasons.
Their participation does not depend on the price innovation
or the spread, and their direction is independent of the innovation:
\[
    D_i\perp Z_h
    \quad\text{conditional on the order being uninformed}.
\]
Since $\mathbb{E}[Z_h]=0$, their contribution to expected
LP profit per unit of potential flow is
\[
    \pi_U(h,s)
    =(1-\theta)\mathbb{E}[s-D_i Z_h\mid U]
    =(1-\theta)s
\]

This inelastic participation of uninformed traders makes the financing of informed losses transparent: changing the spread changes the revenue earned on each uninformed order while leaving the amount of such flow fixed. The baseline requires independence of uninformed direction and the innovation, but does not require buy and sell directions to be equally likely. Section \ref{sec:model-drift} adds directional balance when the reference has nonzero drift.

Informed traders observe $Z_h$ before execution and choose
\[D_i=\operatorname{sign}(Z_h)\]
They execute only when the stale quote offers a strictly
positive trading opportunity:
\[|Z_h|>s\]
At equality, we assume they do not trade.
Their contribution to expected LP profit is therefore
\begin{align*}
    \pi_I(h,s)
      &=\theta\,
        \mathbb{E}\!\left[
          (s-|Z_h|)\mathbf{1}_{\{|Z_h|>s\}}
        \right]\\
      &=-\theta\,\mathbb{E}\!\left[(|Z_h|-s)_+\right],
\end{align*}
where $(x)_+:=\max\{x,0\}$. In what follows, define
\[A_h(s):=\mathbb{E}\!\left[(|Z_h|-s)_+\right]\]

The function $A_h(s)$ measures expected loss per unit of potential informed notional before screening. It combines the probability of a profitable execution with the expected loss conditional on that execution. Movements within the spread generate no informed trade, while movements beyond it generate a loss equal to the excess. This participation rule gives posted quotes their option-like character \citep{copelandgalai1983}.

For an AMM, this loss has the same economic source as arbitrage against stale marginal prices. Its measurement differs from loss-versus-rebalancing, which accounts for changes in pool holdings as arbitrageurs trade along the liquidity curve \citep{milionis2022lvr}. Here, order size is fixed and small, and $A_h(s)$ prices the execution opportunity attached to that order.
Expected LP profit per unit of potential flow is
\[ \pi(h,s,\theta)=(1-\theta)s-\theta A_h(s)\]
Expected profit in a single interval is $h\pi(h,s,\theta)$.
Dividing by its duration gives expected profit per unit of time
$\pi(h,s,\theta)$, so the normalization does not affect break-even
spreads.

\subsection{Screening}

We summarize order selectivity by an admission rate conditional on trader type. Information about an order or its counterparty can support selective internalization and segmentation of flow, as in \citet{battalioholden2001}. Let $\eta\in[0,1]$ denote the fraction of informed opportunities rejected. This parameter is a reduced-form description of the resulting selection. The baseline assumes that screening is costless, preserves every uninformed order, and rejects informed opportunities independently of the realized price innovation.

Formally, let $B_i$ indicate whether an order is admitted. Then,
\[\mathbb{P}(B_i=1\mid U,Z_h)=1, \quad \mathbb{P}(B_i=1\mid I,Z_h)=1-\eta\]
The case $\eta=0$ represents no screening, while $\eta=1$
represents perfect rejection of informed flow. Admission and execution are distinct. An admitted informed
order executes only if $|Z_h|>s$, so
\[
    \mathbb{P}(\text{execute}\mid I)
      =(1-\eta)\mathbb{P}(|Z_h|>s)
\]
The independence assumption on screening implies
\[
    \pi(h,s,\theta,\eta)
      =(1-\theta)s
       -\theta(1-\eta)A_h(s)
\]
Under these assumptions, rejecting a fraction $\eta$ of informed opportunities removes the same fraction of expected informed losses at a fixed spread. The screening rule changes the number of admitted opportunities without changing their conditional severity. Appendix \ref{app:screening-errors} allows rejection of uninformed orders as well and identifies the relative admission rate that governs the spread. A classifier that preferentially rejects the most profitable informed opportunities would require a loss-weighted description of screening.

\subsection{Competitive break-even spreads}

We use zero expected trading profit as the competitive pricing condition, following the adverse-selection logic of \citet{glostenmilgrom1985}. Expected revenue from uninformed orders must finance expected losses to admitted informed orders. Competition is restricted to symmetric quotes around the specified center, so the condition determines a single half-spread. It is an aggregate pricing condition across both trade directions; the model does not separately optimize the bid and ask or derive the allocation of flow across venues. The competitive break-even half-spread $s^*(h,\theta,\eta)$ satisfies
\[ (1-\theta)s^*(h,\theta,\eta) = \theta(1-\eta)A_h\!\left(s^*(h,\theta,\eta)\right)\]
Equivalently,
\[s^*(h,\theta,\eta) = \frac{\theta(1-\eta)}{1-\theta} A_h\!\left(s^*(h,\theta,\eta)\right)\]

\noindent The venue benchmarks are
\[s_0^*=s^*(h_0,\theta,0), \enspace s_C^*=s^*(h_C,\theta,0), \enspace s_P^*=s^*(h_P,\theta,\eta)\]
The venue labels refer to combinations of the model's primitives. The reference layer and slower unscreened benchmark differ in exposure, while the proprietary maker may differ in both exposure and screening. Assigning a real mechanism to one of these cases requires evidence about its effective quote age and admission rules. The comparison $h_P\geq h_0$ considers when stronger screening can compensate a proprietary maker for exposure at least as long as that of the reference layer.

\subsection{Expected executed volume}

Expected executed volume, expressed as a fraction of potential
volume, is
\[ V(h,s,\theta,\eta)=(1-\theta)+\theta(1-\eta)\mathbb{P}(|Z_h|>s)\]
Without screening, this reduces to
\[ V(h,s,\theta,0)=(1-\theta)+\theta\mathbb{P}(|Z_h|>s)\]

Executed volume combines the fixed uninformed component with the informed opportunities that both survive screening and exceed the spread. A change in volume within this baseline therefore reflects informed admission and participation. At a common spread, two venues may incur the same aggregate informed loss while executing different quantities because the loss per informed execution differs. Assessing the participation or welfare of liquidity traders would require an additional demand or routing mechanism.

For clarity, the informed share of executed notional is
\[\theta_{\mathrm{exec}}(h,s,\theta,\eta)
=\frac{\theta(1-\eta)\mathbb{P}(|Z_h|>s)}{(1-\theta)+\theta(1-\eta)\mathbb{P}(|Z_h|>s)}\]
This distinguishes the potential-flow parameter $\theta$ from an informed share measured among completed trades.

\subsection{Analysis}

Write $r(\theta,\eta):=\theta(1-\eta)/(1-\theta)$ for the ratio
of admitted potential informed flow to uninformed flow, before
informed traders decide whether to execute. The break-even condition is
\begin{equation}
    s^*=r(\theta,\eta)A_h(s^*)
    \tag{BE}
    \label{eq:spread-break-even}
\end{equation}

We first establish that the pricing condition has a unique solution and then vary one primitive at a time. The loss function $A_h$ records the opportunities available to that informed flow over horizon $h$. This separation makes it possible to distinguish changes in flow composition from changes in the distribution of stale-price losses. Proofs are collected in Appendix \ref{app:model-proofs}, with the proof of Theorem \ref{prop:screening-threshold} given in the main text.

\begin{proposition}[Existence and uniqueness]
\label{prop:spread-existence}
For every integer $h\geq 1$, $\theta\in(0,1)$, and
$\eta\in[0,1]$, equation~\eqref{eq:spread-break-even}
has a unique nonnegative solution $s^*(h,\theta,\eta)$.
The solution is strictly positive when $\eta<1$, while
\[
    s^*(h,\theta,1)=0
\]
\end{proposition}

The unique solution defines the spread required to finance informed losses under each trading arrangement. A positive spread is necessary whenever some informed opportunities remain admissible. Perfect screening reduces this component of the spread to zero because the model excludes all other costs of liquidity provision. Having established a well-defined pricing benchmark, we next ask how it changes when a larger share of potential flow is informed.

\begin{proposition}[Potential-flow toxicity]
\label{prop:spread-toxicity}
For every integer $h\geq 1$ and fixed $\eta\in[0,1)$,
the equilibrium spread $s^*(h,\theta,\eta)$ is strictly
increasing in $\theta\in(0,1)$.
When $\eta=1$, the equilibrium spread is zero for every
$\theta\in(0,1)$.
\end{proposition}

Greater potential-flow toxicity widens the required spread through two effects: more opportunities can impose informed losses, and fewer uninformed orders contribute revenue. The wider spread deters some informed executions, but that participation response does not overturn the increase in the break-even spread. This comparison holds exposure and screening fixed. We next hold potential composition fixed and allow the maker to change the composition of admitted flow through screening.

\begin{proposition}[Screening]
\label{prop:spread-screening}
For every integer $h\geq 1$ and fixed $\theta\in(0,1)$,
the equilibrium spread $s^*(h,\theta,\eta)$ is strictly
decreasing in $\eta\in[0,1]$.
\end{proposition}

Screening lowers the spread by removing informed losses while preserving uninformed revenue. At equal exposure horizons, any positive screening rate therefore gives the screened maker a strict spread advantage. The result describes the benefit of the specified admission technology; its strength depends on the assumption that uninformed flow is preserved. The next proposition studies the other means of protection available in the model: shortening the period during which the quote can become stale.

\begin{proposition}[Execution interval]
\label{prop:spread-cadence}
For fixed $\theta\in(0,1)$ and $\eta\in[0,1]$,
the equilibrium spread is nondecreasing in the execution
interval:
\[
    s^*(h+1,\theta,\eta)
    \geq
    s^*(h,\theta,\eta),
    \quad h=1,2,\ldots
\]
Under the continuous-time specification with $\sigma>0$,
the inequality is strict whenever $\eta<1$.
\end{proposition}

Longer exposure adds independent mean-zero innovations to the pricing error. Because the informed-loss payoff is convex in that error, the additional uncertainty weakly raises expected loss at every fixed spread. This argument orders the relevant expected losses without requiring a variance or a Gaussian return distribution. It establishes a weak spread advantage for the faster unscreened venue under the maintained resetting rule.

The comparison holds potential flow composition and screening fixed. Broader changes in market speed can also change which counterparties reach a quote; \citet{menkveldzoican2017} show why the total liquidity effect need not be captured by faster updating alone. AMM models with fees and discrete blocks also connect execution frequency to arbitrage losses while allowing pool prices and arbitrage quantities to adjust \citep{milionis2023fees}. Our result isolates the effect of exposure on the loss attached to a given potential order.

\subsection{Screening versus execution speed}
\label{subsec:screening-threshold}

The previous propositions establish that shorter exposure and stronger screening each reduce the required spread. We now compare venues that possess different amounts of these protections. The question is how much screening a maker with horizon $h_P\geq h_0$ needs to match the spread of an unscreened reference layer. Evaluating both loss functions at the reference spread gives a common point of comparison and yields an exact threshold.
For brevity, we write
\[ s_0:=s_0^*=s^*(h_0,\theta,0)\]
Throughout this comparison, screening retains its maintained interpretation: it is costless, never rejects uninformed orders, and rejects a fraction $\eta$ of informed opportunities independently of their profitability. All spread comparisons refer to the
linearized-profit model.

\begin{theorem}[The exact screening threshold]
\label{prop:screening-threshold}
Define
\begin{equation}
    \eta_{\mathrm{crit}}(h_P,\theta)
    :=1-\frac{A_{h_0}(s_0)}{A_{h_P}(s_0)}
    \label{eq:screening-threshold}
\end{equation}
This threshold is well defined and belongs to $[0,1)$.
The equilibrium spread ranking is characterized exactly by
\begin{equation}
\begin{aligned}
    s^*(h_P,\theta,\eta)>s_0
        &\quad\Longleftrightarrow\quad
        \eta<\eta_{\mathrm{crit}}(h_P,\theta)\\
    s^*(h_P,\theta,\eta)=s_0
        &\quad\Longleftrightarrow\quad
        \eta=\eta_{\mathrm{crit}}(h_P,\theta)\\
    s^*(h_P,\theta,\eta)<s_0
        &\quad\Longleftrightarrow\quad
        \eta>\eta_{\mathrm{crit}}(h_P,\theta)
\end{aligned}
\label{eq:threshold-ranking}
\end{equation}
In particular, $\eta_{\mathrm{crit}}(h_0,\theta)=0$.
For $h_P>h_0$, the threshold is strictly positive whenever
$A_{h_P}(s_0)>A_{h_0}(s_0)$; this condition holds under the
continuous-time specification with $\sigma>0$.
\end{theorem}
\begin{proof}
By the break-even condition of the reference layer,
$A_{h_0}(s_0)=(1-\theta)s_0/\theta>0$.
The function $x\mapsto(|x|-s_0)_+$ is convex. Adding independent
mean-zero increments and applying conditional Jensen's inequality
therefore gives $A_{h_P}(s_0)\geq A_{h_0}(s_0)$.
Thus the threshold is well defined and belongs to $[0,1)$.
At the benchmark spread,
\begin{align*}
    \pi(h_P,s_0,\theta,\eta)
    &=\theta\bigl[A_{h_0}(s_0)-(1-\eta)A_{h_P}(s_0)\bigr]\\
    &=\theta A_{h_P}(s_0)
        [\eta-\eta_{\mathrm{crit}}(h_P,\theta)]
\end{align*}
Profit is strictly increasing in the spread and has a unique zero.
Its sign at $s_0$ therefore gives all three rankings.
The equal-horizon case and the condition for a positive threshold
follow directly from the defining ratio.
\end{proof}

At the reference spread $s_0$, the slower maker would incur the unscreened loss $A_{h_P}(s_0)$, whereas the reference layer incurs $A_{h_0}(s_0)$. The threshold is the fraction of the former that must be removed to make these losses equal. Both venues then earn the same revenue from the common uninformed flow and can support the same spread. Evaluating losses at this common spread is essential because each venue's own break-even spread already incorporates its response to adverse selection.

The equivalence between a rejection rate and a loss-reduction rate follows from screening being independent of opportunity size. The theorem therefore gives a technological comparison under a common potential-flow composition. Its threshold can be used to assess whether an assumed screening capability offsets a given exposure disadvantage; allocating orders between venues requires a further routing model.

\begin{proposition}[Horizon, scale, and the low-toxicity limit]
\label{prop:threshold-robust-comparisons}
The screening threshold has the following properties.
\begin{enumerate}
    \item For every $\theta\in(0,1)$,
    \[
        \eta_{\mathrm{crit}}(h_P+1,\theta)
        \geq\eta_{\mathrm{crit}}(h_P,\theta)
    \]
    The inequality is strict under the continuous-time specification
    with $\sigma>0$.

    \item Multiplying every price innovation by the same constant
    $c>0$ multiplies every equilibrium spread by $c$ and leaves
    $\eta_{\mathrm{crit}}(h_P,\theta)$ unchanged.

    \item For each fixed $h_P$, the threshold is continuous in
    $\theta\in(0,1)$ and satisfies
    \begin{equation}
        \lim_{\theta\downarrow0}
        \eta_{\mathrm{crit}}(h_P,\theta)
        =1-\frac{\mathbb E[|Z_{h_0}|]}{\mathbb E[|Z_{h_P}|]}
        \label{eq:threshold-low-toxicity}
    \end{equation}
\end{enumerate}
\end{proposition}

The threshold separates the scale of execution costs from the relative effectiveness of screening and speed. A longer exposure weakly increases the screening required to match the reference spread. Multiplying the entire innovation process by a common positive scale raises both venues' spreads proportionally and leaves the threshold unchanged. Changes in tail shape, jump intensity, or only one component of the innovation process need not have this property.

As informed opportunities become rare, both spreads approach zero, and the threshold converges to a ratio of expected absolute innovations. A positive limiting threshold can therefore coexist with a very small absolute spread difference. Whether the threshold rises with toxicity away from this limit requires a different comparison: the severity of the informed orders that remain profitable at the benchmark spread.

\begin{proposition}[Toxicity, tail shape, and volume at the threshold]
\label{prop:threshold-toxicity-volume}
Suppose
\[\mathbb{P}(|Z_{h_0}|=s_0)=\mathbb{P}(|Z_{h_P}|=s_0)=0\]
Then the threshold is differentiable with respect to $\theta$ at
the parameter value under consideration, and
\begin{equation}
\begin{aligned}
    &\frac{\partial\eta_{\mathrm{crit}}(h_P,\theta)}{\partial\theta}\\
    &\quad={}\frac{s_0[1-\eta_{\mathrm{crit}}(h_P,\theta)]}
    {\theta[1-\theta+\theta\mathbb{P}(|Z_{h_0}|>s_0)]}\\
    &\qquad\times\left[
        \frac{\mathbb{P}(|Z_{h_0}|>s_0)}{A_{h_0}(s_0)}
        -\frac{\mathbb{P}(|Z_{h_P}|>s_0)}{A_{h_P}(s_0)}
    \right]
\end{aligned}
\label{eq:threshold-toxicity-derivative}
\end{equation}
Without any atomlessness assumption, executed volumes at the
spread-matching screening rate satisfy
\begin{equation}
\begin{aligned}
    &V\bigl(h_P,s_0,\theta,\eta_{\mathrm{crit}}(h_P,\theta)\bigr)
        -V(h_0,s_0,\theta,0)\\
    &\quad=\theta A_{h_0}(s_0)
    \left[
        \frac{\mathbb{P}(|Z_{h_P}|>s_0)}{A_{h_P}(s_0)}
        -\frac{\mathbb{P}(|Z_{h_0}|>s_0)}{A_{h_0}(s_0)}
    \right]
\end{aligned}
\label{eq:threshold-volume-difference}
\end{equation}
Consequently, the derivative in
\eqref{eq:threshold-toxicity-derivative} is positive exactly when
the screened maker executes less volume than the reference layer at their common spread.
The maintained assumptions alone imply neither a monotone threshold
in $\theta$ nor a universal volume ranking at the threshold.
\end{proposition}

Greater toxicity raises each incompletely screened venue's spread, but it also changes which informed opportunities remain profitable at the reference spread. The resulting threshold comparison depends on expected loss per executed informed order, measured by $A_h(s_0)/\mathbb{P}(|Z_h|>s_0)$. Convexity orders aggregate losses across horizons; it does not by itself order this conditional severity. The threshold's response to toxicity can consequently have either sign under the general innovation assumptions.

The same conditional severity explains volume at the matching threshold. Both venues must finance the same aggregate informed loss, $(1-\theta)s_0$. If the slower maker's surviving informed executions are more costly on average, fewer of them can be admitted while maintaining that equality. The Gaussian case below supplies the additional tail ordering that makes this comparison unambiguous.

\begin{proposition}[Gaussian benchmark]
\label{prop:threshold-gaussian}
Suppose $dp_t=\sigma\,dW_t$ with $\sigma>0$ and no jump component.
Let $\phi$ and $\Phi$ denote the standard normal density and
distribution function. Then
\begin{equation}
    A_h(s)=2\left[
        \sigma\sqrt h\,\phi\!\left(\frac{s}{\sigma\sqrt h}\right)
        -s\left\{1-\Phi\!\left(\frac{s}{\sigma\sqrt h}\right)\right\}
    \right],
    \label{eq:gaussian-adverse-selection}
\end{equation}
and
\begin{equation}
    s^*(h,\theta,\eta)=\sqrt{h/h_0}\,s^*(h_0,\theta,\eta)
    \label{eq:gaussian-spread-scaling}
\end{equation}
For every $h_P>h_0$, $\eta_{\mathrm{crit}}(h_P,\theta)$ is
independent of $\sigma$, strictly increasing in $\theta$, and obeys
\begin{equation}
    \lim_{\theta\downarrow0}\eta_{\mathrm{crit}}(h_P,\theta)
        =1-\sqrt{\frac{h_0}{h_P}},
    \quad
    \lim_{\theta\uparrow1}\eta_{\mathrm{crit}}(h_P,\theta)=1
    \label{eq:gaussian-threshold-limits}
\end{equation}
At the spread-matching screening rate, the longer-horizon maker
executes strictly less volume than the reference layer:
\[V\bigl(h_P,s_0,\theta,\eta_{\mathrm{crit}}(h_P,\theta)\bigr)<V(h_0,s_0,\theta,0)\]
\end{proposition}

The Gaussian benchmark yields a transparent scale comparison: for fixed toxicity and screening, the break-even spread grows with the square root of the exposure horizon. This follows from Brownian scaling and the homogeneity of the pricing equation. Gaussian tails also imply that the longer horizon has greater loss per profitable informed execution at a common spread. As a result, the matching screening threshold increases with toxicity, and the screened maker executes less informed volume at that threshold.

These sharper rankings depend on the Gaussian specification. The general results above remain applicable when the innovation distribution has jumps or atoms, while the Gaussian case provides an analytical benchmark for interpreting horizon ratios. We next consider equal horizons, where the direction of the screening advantage is already known, and its magnitude becomes the relevant object.

\begin{proposition}[The equal-cadence spread premium]
\label{prop:equal-cadence-premium}
Define
\[
    \Delta s(\theta,\eta)
        :=s_0-s^*(h_0,\theta,\eta)
\]
For fixed $\theta\in(0,1)$, the premium is continuous and
strictly increasing in $\eta\in[0,1]$, with
\[
    \Delta s(\theta,0)=0,
    \quad
    \Delta s(\theta,1)=s_0
\]
For every $\eta\in[0,1]$, it satisfies the exact identity
\begin{equation}
    \Delta s(\theta,\eta)
    +r(\theta,\eta)
       \int_{s^*(h_0,\theta,\eta)}^{s_0}
           \mathbb{P}(|Z_{h_0}|>u)\,du
    =\eta s_0
    \label{eq:equal-cadence-premium}
\end{equation}
In particular, $0<\Delta s(\theta,\eta)<\eta s_0$ for $0<\eta<1$.
As screening is introduced from zero,
\begin{equation}
    \Delta s(\theta,\eta)
      =\eta\,
       \frac{(1-\theta)s_0}
       {1-\theta+\theta\mathbb{P}(|Z_{h_0}|\geq s_0)}
       +o(\eta),
    \quad \eta\downarrow0
    \label{eq:equal-cadence-premium-local}
\end{equation}
The weak inequality inside the probability makes this expansion
valid even when $|Z_{h_0}|$ has an atom at $s_0$.
\end{proposition}

The quantity $\Delta s$ is the unscreened venue's half-spread premium at a common exposure horizon. Screening initially removes informed opportunities, but the resulting narrower spread makes additional opportunities profitable. This participation response partly offsets the direct benefit, giving $\Delta s<\eta s_0$ for an interior screening rate. The local expansion also accounts for traders who are indifferent at the original spread and begin trading as the spread falls.

This premium connects the model to comparisons of execution quality at equal speed. Exact equality requires zero effective screening under the maintained primitives; practical proximity can be assessed by expressing $\Delta s$ in basis points and specifying a tolerance.

\section{Predictable Drift and Stale-Quote Exposure}
\label{sec:model-drift}

We now allow a known predictable component in log returns and distinguish its effect on the quote's center from its effect on the required half-spread. We begin with the restriction that the quote remains centered on the current reference throughout the interval. Under this restriction, predictable displacement contributes to the stale-price opportunities available at execution. Proposition \ref{prop:drift-recentering} then allows the maker to incorporate the known drift into the initial quote center and identifies which spread effects disappear. This sequence makes the role of the centering restriction explicit. Proofs are collected in Appendix \ref{app:drift-proofs}.

\subsection{The same execution model with nonzero drift}

Replace the zero-mean increment assumption with
\[
    \varepsilon_t=\mu+\xi_t,\quad
    \xi_t\ \text{i.i.d.},\quad
    \mathbb E[\xi_t]=0,\quad
    0<\mathbb E[|\xi_t|]<\infty
\]
The centered increments are independent of information available
before the interval. The constant $\mu\in\mathbb R$ is known when
the quote is set. Define
\[
    M_{t,h}:=\sum_{k=0}^{h-1}\xi_{t+k},
    \quad
    Z_{t,h}(\mu):=p_{t+h}-p_t=\mu h+M_{t,h}
\]
As before, stationarity allows us to suppress $t$.
Thus $\mathbb E[Z_h(\mu)]=\mu h$. The case $\mu=0$ recovers the benchmark. A nonzero constant mean is not an assumption of serial correlation or momentum in the centered innovations.

Initially, quotes remain centered at $p_t$, not at the expected terminal log price $p_t+\mu h$. This is the same centering restriction as in the benchmark. Its relaxation is considered separately below. Informed traders therefore choose the sign of $Z_h(\mu)$ and execute if and only if $|Z_h(\mu)|>s$.

Independence of uninformed direction and the innovation now gives
\[
    \mathbb E[s-D_iZ_h(\mu)\mid U]
      =s-\mu h\,\mathbb E[D_i\mid U]
\]
To retain the original uninformed-flow contribution, impose
\begin{equation}
    \mathbb E[D_i\mid U]=0
    \label{eq:drift-balanced-flow}
\end{equation}
The assumption concerns the average initial notional; individual intervals may still contain unequal buys and sells. It isolates the spread needed to finance informed losses from gains or losses created by directional liquidity demand.

Define the expected informed loss and the participation probability by
\begin{gather*}
    A_h(s;\mu):=\mathbb E[(|Z_h(\mu)|-s)_+],\\
    S_h(s;\mu):=\mathbb P(|Z_h(\mu)|>s)
\end{gather*}
Under \eqref{eq:drift-balanced-flow}, profit and volume are
\begin{align}
    \pi(h,s,\theta,\eta;\mu)
       &=(1-\theta)s-\theta(1-\eta)A_h(s;\mu),
       \label{eq:drift-profit}\\
    V(h,s,\theta,\eta;\mu)
       &=(1-\theta)+\theta(1-\eta)S_h(s;\mu)
       \label{eq:drift-volume}
\end{align}
The competitive spread solves
\begin{equation}
\begin{aligned}
        s^*(h,\theta,\eta;\mu)
       &=r(\theta,\eta)
          A_h\bigl(s^*(h,\theta,\eta;\mu);\mu\bigr),
    \quad\\
    r(\theta,\eta)&=\frac{\theta(1-\eta)}{1-\theta}
\end{aligned}
    \label{eq:drift-break-even}
\end{equation}
The added argument records drift, and all other notation and the interpretation of screening are unchanged.

\begin{proposition}[Equilibrium and screening with drift]
\label{prop:drift-equilibrium}
For every integer $h\geq1$, $\theta\in(0,1)$, $\eta\in[0,1]$,
and $\mu\in\mathbb R$, equation~\eqref{eq:drift-break-even} has a
unique nonnegative solution. It is positive for $\eta<1$ and zero
for $\eta=1$, and is jointly continuous in $(\theta,\eta,\mu)$.
For fixed $h$ and $\mu$, it is strictly increasing in $\theta$
when $\eta<1$, and strictly decreasing in $\eta$.
\end{proposition}

Once expected uninformed demand is balanced, drift changes the informed-loss function while leaving the break-even argument intact. Existence, uniqueness, and the comparative statics in toxicity and screening therefore extend without imposing symmetry on the innovation law. Ordering losses across different exposure horizons requires an additional argument, because accumulating drift can interact with asymmetry in the unpredictable component.

\subsection{Execution speed and the screening threshold}

Comparing different horizons requires more structure. For the
results in this subsection, assume that the centered increments
are symmetric:
\begin{equation}
    \xi_t\overset{d}{=}-\xi_t
    \label{eq:drift-symmetry}
\end{equation}
This condition permits atoms, jumps, and infinite variance.
One continuous-time specification is
\[
    dp_t=\mu\,dt+\sigma\,dW_t+d\widetilde J_t,
\]
where $\widetilde J_t$ is an independent, integrable, symmetric
Lévy jump process. Its increments have mean zero.

Symmetry ensures that a displacement of the quote center has the same expected cost in either direction. Combined with convexity of the loss payoff, it makes a larger absolute displacement weakly more costly. This condition supplies the ordering needed when a longer horizon increases both accumulated drift and residual uncertainty.

\begin{proposition}[Cadence and the magnitude of drift]
\label{prop:drift-cadence}
Under \eqref{eq:drift-symmetry}, for every $s\geq0$ the loss
$A_h(s;\mu)$ is nondecreasing in the integer horizon $h$, is even
in $\mu$, and is nondecreasing in $|\mu|$. Consequently,
\[ s^*(h+1,\theta,\eta;\mu)\geq s^*(h,\theta,\eta;\mu),\]
and the equilibrium spread is even in $\mu$ and nondecreasing in its absolute value.

Under the continuous-time specification with $\sigma>0$ and $\eta<1$, the horizon comparison is strict, and any strict increase in $|\mu|$ strictly increases the spread.
\end{proposition}

For a spot-centered quote with balanced liquidity demand and symmetric residual innovations, upward and downward drift have the same spread cost. Longer exposure accumulates a larger predictable displacement and more unpredictable variation. The symmetry restriction is substantive: without it, drift may move probability mass toward the fixed quote center and reverse the horizon comparison, as the counterexample in Appendix \ref{app:drift-counterexample} shows. With loss ordering established, the screening threshold follows from the same common-spread comparison as in the baseline.

Fix the same benchmark horizon $h_0$ as before and write
\[s_0(\mu):=s^*(h_0,\theta,0;\mu)\]

\begin{proposition}[The screening threshold with drift]
\label{prop:drift-threshold}
Under \eqref{eq:drift-symmetry}, for every integer $h_P\geq h_0$,
\begin{equation}
    \eta_{\mathrm{crit}}(h_P,\theta;\mu)
      :=1-\frac{A_{h_0}(s_0(\mu);\mu)}
                    {A_{h_P}(s_0(\mu);\mu)}
      \in[0,1)
    \label{eq:drift-threshold}
\end{equation}
The complete spread ranking is
\[
\begin{aligned}
    s^*(h_P,\theta,\eta;\mu)>s_0(\mu)
       &\ \Longleftrightarrow\
         \eta<\eta_{\mathrm{crit}}(h_P,\theta;\mu)\\
    s^*(h_P,\theta,\eta;\mu)=s_0(\mu)
       &\ \Longleftrightarrow\
         \eta=\eta_{\mathrm{crit}}(h_P,\theta;\mu)\\
    s^*(h_P,\theta,\eta;\mu)<s_0(\mu)
       &\ \Longleftrightarrow\
         \eta>\eta_{\mathrm{crit}}(h_P,\theta;\mu)
\end{aligned}
\]
The threshold is zero when $h_P=h_0$, is nondecreasing in $h_P$,
and is continuous in $(\theta,\mu)$. If $\sigma>0$ in the
continuous-time specification, it is strictly increasing in $h_P$.
Moreover,
\begin{equation}
    \lim_{\theta\downarrow0}
      \eta_{\mathrm{crit}}(h_P,\theta;\mu)
       =1-\frac{\mathbb E|Z_{h_0}(\mu)|}
                    {\mathbb E|Z_{h_P}(\mu)|}
    \label{eq:drift-low-toxicity}
\end{equation}
Multiplying both $\mu$ and every centered innovation by the same
positive constant multiplies all spreads by that constant and
leaves the threshold unchanged.
\end{proposition}

The threshold again measures the share of the slower maker's loss that must be removed to support the reference spread. Drift changes both that spread and the losses evaluated there, so it enters the comparison through both objects. Scale invariance now requires a common rescaling of the predictable and unpredictable components. Changing volatility while keeping drift fixed changes their relative importance and can therefore change the threshold.

\begin{proposition}[Toxicity, loss severity, and the equal-cadence premium]
\label{prop:drift-identities}
Maintain \eqref{eq:drift-symmetry}. At a parameter value such that
$\mathbb P(|Z_{h_0}(\mu)|=s_0(\mu))
=\mathbb P(|Z_{h_P}(\mu)|=s_0(\mu))=0$, the threshold derivative is
\begin{equation}
\begin{aligned}
    \frac{\partial\eta_{\mathrm{crit}}}{\partial\theta}
    ={}&
    \frac{s_0(\mu)(1-\eta_{\mathrm{crit}})}
         {\theta[1-\theta+\theta S_{h_0}(s_0(\mu);\mu)]}\\
    &\times
    \left[
      \frac{S_{h_0}(s_0(\mu);\mu)}{A_{h_0}(s_0(\mu);\mu)}
      -\frac{S_{h_P}(s_0(\mu);\mu)}{A_{h_P}(s_0(\mu);\mu)}
    \right],
\end{aligned}
\label{eq:drift-toxicity-derivative}
\end{equation}
where $\eta_{\mathrm{crit}}=\eta_{\mathrm{crit}}(h_P,\theta;\mu)$.
Without an atomlessness assumption, the volume at the matching spread
satisfies
\begin{equation}
\begin{aligned}
    &V(h_P,s_0(\mu),\theta,\eta_{\mathrm{crit}};\mu)
       -V(h_0,s_0(\mu),\theta,0;\mu)=\\
    &\theta A_{h_0}(s_0(\mu);\mu)
    \left[
      \frac{S_{h_P}(s_0(\mu);\mu)}{A_{h_P}(s_0(\mu);\mu)}
      -\frac{S_{h_0}(s_0(\mu);\mu)}{A_{h_0}(s_0(\mu);\mu)}
    \right]
\end{aligned}
\label{eq:drift-volume-comparison}
\end{equation}
Thus, the same expected loss per executed informed order governs the toxicity derivative and the opposite volume ranking.

For any fixed $h$, define the equal-cadence premium by
\[\Delta_h(\theta,\eta;\mu):=s^*(h,\theta,0;\mu)-s^*(h,\theta,\eta;\mu)\]
It is continuous and strictly increasing in $\eta$, from zero at $\eta=0$ to $s^*(h,\theta,0;\mu)$ at $\eta=1$, and
\begin{equation}
\begin{aligned}
        \Delta_h+{}&
      r(\theta,\eta)
       \int_{s^*(h,\theta,\eta;\mu)}^{s^*(h,\theta,0;\mu)}
                 S_h(u;\mu)\,du\\
      &=\eta s^*(h,\theta,0;\mu)
\end{aligned}
    \label{eq:drift-premium}
\end{equation}
For $0<\eta<1$, its value lies strictly between zero and
$\eta s^*(h,\theta,0;\mu)$. Writing $s_h^{\mathrm{ns}}:=s^*(h,\theta,0;\mu)$
for the unscreened spread,
\[
    \Delta_h(\theta,\eta;\mu)
       =\eta\,
          \frac{(1-\theta)s_h^{\mathrm{ns}}}
          {1-\theta+\theta\mathbb P(|Z_h(\mu)|\geq s_h^{\mathrm{ns}})}
          +o(\eta),
          \quad \eta\downarrow0
\]
The premium statements do not require symmetry.
\end{proposition}

The distinction between total informed loss and loss per informed execution continues to govern the comparison. Symmetry and drift alone do not fix the sign of the threshold's toxicity derivative or the volume ranking at a matched spread. At a common horizon, screening still lowers the spread, with part of its benefit offset by additional informed participation. The Gaussian specification below provides the stronger tail restriction needed to recover definite toxicity and volume rankings across horizons.

\subsection{Gaussian innovations with drift}

\noindent Suppose now that
\[dp_t=\mu\,dt+\sigma\,dW_t,\quad \sigma>0\]
Then $Z_h(\mu)$ is normal with mean $\mu h$ and variance $\sigma^2h$.
The parameter $\mu$ is the drift of the \emph{log} price. It is not
the arithmetic-price drift: Itô's formula gives
$dP_t/P_t=(\mu+\sigma^2/2)\,dt+\sigma\,dW_t$.

\begin{proposition}[Gaussian loss and threshold comparisons]
\label{prop:drift-gaussian}
Let $\overline\Phi=1-\Phi$. The informed-loss function is
\begin{equation}
\begin{aligned}
 &A_h(s;\mu)={}
 \sigma\sqrt h\left[
       \phi\!\left(\frac{s-\mu h}{\sigma\sqrt h}\right)
      +\phi\!\left(\frac{s+\mu h}{\sigma\sqrt h}\right)\right]\\
 &+(\mu h-s)\overline\Phi\!\left(\frac{s-\mu h}{\sigma\sqrt h}\right)
   +(-\mu h-s)\overline\Phi\!\left(\frac{s+\mu h}{\sigma\sqrt h}\right)
\end{aligned}
\label{eq:drift-gaussian-loss}
\end{equation}
For every $h_P>h_0$ and every fixed $\mu$, the screening threshold
is strictly increasing in $\theta$, and
\[
    \lim_{\theta\uparrow1}
          \eta_{\mathrm{crit}}(h_P,\theta;\mu)=1
\]
At the matching spread, the slower screened maker executes strictly
less volume than the unscreened reference layer. Write
\[
    \kappa_h(\mu)
       :=2\sigma\sqrt h\,\phi\!\left(\frac{|\mu|\sqrt h}{\sigma}\right)
          +|\mu|h\left[
                 2\Phi\!\left(\frac{|\mu|\sqrt h}{\sigma}\right)-1
                \right]
\]
Then the low-toxicity threshold is
\begin{equation}
    \lim_{\theta\downarrow0}
          \eta_{\mathrm{crit}}(h_P,\theta;\mu)
       =1-\frac{\kappa_{h_0}(\mu)}{\kappa_{h_P}(\mu)}
    \label{eq:drift-gaussian-low-toxicity}
\end{equation}
For $h_P>h_0$, this limit is strictly increasing in $|\mu|/\sigma$.
It equals $1-\sqrt{h_0/h_P}$ at $\mu=0$ and converges to
$1-h_0/h_P$ as $|\mu|/\sigma\to\infty$.

More generally, the threshold depends on the Gaussian price
parameters through $|\mu|\sqrt{h_0}/\sigma$, as well as on $\theta$
and $h_P/h_0$. The square-root spread scaling of
Proposition \ref{prop:threshold-gaussian} applies at $\mu=0$;
for $\eta<1$ it does not extend to fixed nonzero $\mu$.
\end{proposition}

With Gaussian innovations, the longer exposure has greater expected loss per profitable informed execution at any common spread, even when drift is present. Matching aggregate losses consequently requires fewer informed executions, and the required screening rate rises with toxicity. The dimensionless drift magnitude $|\mu|\sqrt{h_0}/\sigma$ measures predictable displacement relative to diffusion uncertainty at the reference horizon; together with toxicity and the horizon ratio, it determines the threshold.

The low-toxicity limit illustrates the change in time scaling: predictable displacement accumulates linearly with the horizon, while diffusion uncertainty grows with its square root. Its monotonicity in relative drift does not establish monotonicity of the threshold in drift at every positive toxicity. The next two results instead characterize the spread effect near zero drift and when residual uncertainty vanishes.

\begin{proposition}[The local spread effect of drift]
\label{prop:drift-local}
Under Gaussian innovations, fix $h$, $\theta\in(0,1)$,
$\eta<1$, and $\sigma>0$. Write
$s^{\mathrm{nd}}:=s^*(h,\theta,\eta;0)$ for the no-drift spread and $r=r(\theta,\eta)$. Then, as $\mu\to0$,
\begin{equation}
\begin{aligned}
    &s^*(h,\theta,\eta;\mu)\\
      &=s^{\mathrm{nd}}
       +\frac{r h^{3/2}
               \phi\!\left(s^{\mathrm{nd}}/(\sigma\sqrt h)\right)}
              {\sigma[1+rS_h(s^{\mathrm{nd}};0)]}\,\mu^2
       +o(\mu^2)
\end{aligned}
\label{eq:drift-local-expansion}
\end{equation}
In particular, the first derivative at zero is zero, and the
quadratic coefficient is positive. For $\eta=1$, the spread
is identically zero.
\end{proposition}

Around zero drift, the Gaussian loss is smooth and symmetric in the direction of displacement, so the spread has no first-order response. Its leading response is positive and quadratic in the uncorrected drift. This local result describes sensitivity near the benchmark; assessing its size requires comparing predictable displacement with the uncertainty accumulated over the relevant horizon. Distributions with atoms and directional uninformed demand need not have the same expansion.

\begin{proposition}[The drift-dominated limit]
\label{prop:drift-dominated}
Return to integrable centered innovations and fix $\mu\ne0$.
Replace $\xi_t$ by $c\xi_t$, where $c>0$, and denote the resulting
spreads and threshold by $s_c^*$ and $\eta_{\mathrm{crit},c}$.
Then, as $c\downarrow0$,
\[
    s_c^*(h,\theta,\eta;\mu)
       \longrightarrow
       \frac{\theta(1-\eta)}{1-\theta\eta}\,|\mu|h
\]
Under the symmetry assumption, for $h_P\geq h_0$,
\begin{equation}
    \eta_{\mathrm{crit},c}(h_P,\theta;\mu)
       \longrightarrow
       \frac{h_P-h_0}{h_P-\theta h_0}
    \label{eq:drift-dominated-threshold}
\end{equation}
\end{proposition}

As residual uncertainty vanishes, the entire pricing discrepancy comes from the known drift that the fixed center has left uncorrected. The required spread then scales linearly with exposure, and the limiting screening threshold has a closed form. This limit can be taken with small absolute log-price changes, so the contrast with square-root diffusion scaling does not depend on large-return behavior. It also makes the centering restriction especially visible: the maker is financing losses from a movement that was predictable when the quote was set.

\subsection{Allowing the quote center to reflect drift}

We now allow the initial quote center to reflect the same drift that informed traders anticipate. The maker still commits to a fixed quote over the interval, so its ability to respond to subsequent innovations is unchanged. Only the predictable location of the terminal reference is incorporated at the time the quote is set.

\begin{proposition}[Forecast-adjusted quote centers]
\label{prop:drift-recentering}
Suppose the maker may choose, at the beginning of the interval, a fixed log-quote center $p_t+\gamma_h$, so that its bid and ask log prices are $p_t+\gamma_h-s$ and $p_t+\gamma_h+s$. If $\gamma_h=\mu h$, the relevant terminal pricing error is $M_h$. The equilibrium spread is exactly that of the zero-drift model with centered increments $\xi_t$:
\[s_{\mathrm{adj}}^*(h,\theta,\eta;\mu)=s^*(h,\theta,\eta;0)\]
This equality does not require symmetry or balanced uninformed directions; independence of the uninformed direction and $M_h$ suffices.

If uninformed flow is balanced and the centered increments are
symmetric, $\gamma_h=\mu h$ minimizes the informed-loss function over
all fixed quote centers for every given half-spread. It therefore
attains the smallest competitive half-spread available by choosing
the center. Under the continuous-time specification with
$\sigma>0$ and $\eta<1$, this minimizing center is unique.
\end{proposition}

A known additive drift can be incorporated into the quote center, leaving the spread to finance losses from residual innovations. Under symmetric residuals and balanced uninformed flow, this center also minimizes the required half-spread among fixed centers. The earlier drift premium therefore measures the cost of a centering restriction or an uncorrected forecast error. Comparisons across trading arrangements must specify both their quote-exposure horizons and their ability to incorporate predictable price changes before committing the quote.

The result relies on drift entering as a known location shift of an otherwise unchanged innovation law. An estimated or state-dependent forecast can also change the conditional distribution of the residual, including its tails. In that setting, the relevant objects are the expected excess-loss functions of the forecast errors. A reduction in mean squared forecast error alone need not reduce these losses at the spreads being compared.

\section{Empirical Analysis}\label{sec:analysis}

In this section, we take the model developed in Sections \ref{sec:model}-\ref{sec:model-drift} to the data. Throughout, the reference execution layer is the one-second subslot, i.e., $h_0 = 1$\,s, and both the slow unscreened benchmark and the screened proprietary maker quote at the twelve-second slot, i.e., $h_C = h_P = 12$\,s. The objects this section estimates are those the theory delivers: the fundamental shock $\kappa = \mathbb{E} |Z_h| = A_h (0)$ that sets the scale of the adverse-selection function; the participation-constrained half-spread $s^*(\theta)$ of Propositions \ref{prop:spread-existence}-\ref{prop:spread-cadence}; the role of drift in the shock and the direction of the loss, which we measure in the execution price; the growth of the stale-quote discrepancy with elapsed staleness; and the screening threshold $\eta_{\mathrm{crit}}$ of Theorem \ref{prop:screening-threshold}, the rate at which a slow maker matches the reference layer's spread.

Our task is not a simple search for correlations: each measurement is complicated by a specific confounding factor. We cannot simply regress the cost of staleness on cadence because trade timing is chosen and correlated with pool depth and same-block batching; we instead use the variation in staleness that missed proposer slots create. Toxicity cannot be read from the sign of markouts because, against a martingale reference, the favorable-markout share is set by the symmetric price innovation, not by the informed share: the informed tilt is second-order to the diffusion, so the share barely moves with toxicity. Our methodology is therefore built around an exogenous price benchmark and exogenous variation in block timing, which address the first problem, and it treats toxicity as a parameter rather than reading it from markouts, which addresses the second.

\subsection{Methodology}

\runhead{Sample and data} We use the Binance ETH-USDT best bid and ask as the benchmark, reduced to a one-second midpoint $m_t$ (the midpoint of the last best bid and ask in each second) over the period from 1 June to 29 August 2026. On-chain execution is decoded from raw swap logs for four USDC-WETH pools spanning two AMM generations, Uniswap v2 and v3, as shown in Table \ref{tab:summary}. Each swap is timed at its block's slot time, read from the block header, and after removing dust swaps (below \$1 or $10^{-4}$\,ETH), the sample holds 1,725,539 swaps, and the batch-netting statistic uses the full Binance trade tape of 260 million trades (33.4 per second). For a fill $j$ at pool $v$, time $t_j$, taker direction $q_j \in \{-1, +1\}$, and execution price $p_j^{\text{ex}}$, we build the signed markout \eqref{eq:markout}, the effective spread \eqref{eq:effspread}, and the elapsed staleness \eqref{eq:staleness}:
\begin{align}
\mathrm{MO}_j (\tau) &= q_j \frac{m_{t_j + \tau} - m_{t_j}}{m_{t_j}}, \label{eq:markout}\\
\mathrm{ES}_j &= 2q_j \frac{p_j^{\text{ex}} - m_{t_j}}{m_{t_j}}, \label{eq:effspread}\\
a_j &= t_j - t^{\text{last-upd}}_v, \label{eq:staleness}
\end{align}
where $t^{\text{last-upd}}_v$ is the time of the previous swap in pool $v$, and the markout is the taker's realized gain and, hence, the liquidity provider's adverse selection. Both ratios in \eqref{eq:markout} and \eqref{eq:effspread} are multiplied by $10^4$ and reported in basis points throughout. The local drift $\hat{\mu}_t$ is the trailing sixty-second benchmark return, and a second is trending when $|\hat{\mu}_t| > 10$\, bps; by this cut, 88.9\% of seconds are range-bound and 11.1\% trending, and 318,822 of the swaps fall in trending seconds.

\begin{table}[t]
\caption{Summary of the four USDC-WETH pools and their samples.}
\label{tab:summary}
\centering
\begin{tabular*}{\columnwidth}{@{}l@{\extracolsep{\fill}}rr@{}}
\toprule
Pool (fee tier) & Number of swaps & Volume in \$M \\
\midrule
Uniswap v3 (1\,bp) & 1,290,417 & 2,809 \\
Uniswap v3 (5\,bps) & 349,391 & 6,261 \\
Uniswap v2 (30\,bps) & 56,378 & 57 \\
Uniswap v3 (30\,bps) & 29,353 & 259 \\
\bottomrule
\end{tabular*}
\tabnotes{The sample runs from 1 June to 29 August 2026. Volume is the USDC notional of the swaps. Swaps are timed at their block's slot time, and swaps below \$1 or $10^{-4}$\,ETH are excluded. Of the 1,725,539 swaps, 318,822 fall in trending seconds. The benchmark is 7,776,000 one-second observations of the Binance ETH-USDT mid; a quote is observed in 95.5\% of seconds, and the last quote is carried forward otherwise. Uniswap v4 pools are excluded because their swap data end on 30 June 2026.}
\end{table}

\begin{figure*}[t]
\includegraphics[width=\textwidth]{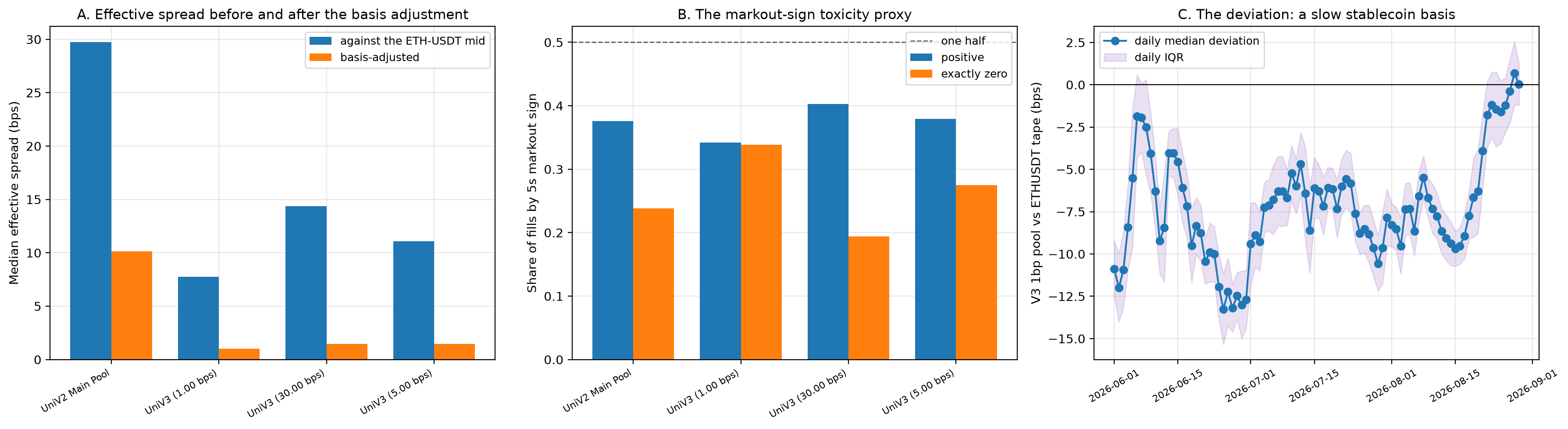}
\caption{Feature-layer construction and integrity. (A) Median effective spread by pool, against the ETH-USDT mid and after subtracting the three-hundred-second rolling median basis. (B) The markout-sign toxicity proxy: the share of fills with a positive five-second markout lies below one-half in every pool because 19-34\% of five-second markouts are exactly zero. (C) Daily median and interquartile range of the deviation of the v3 1\, bp pool from the benchmark: a slow stablecoin basis, not staleness.}
\label{fig:features}
\end{figure*}

\runhead{Feature engineering} Decoded swap flow cannot carry a spread measurement until the three steps are applied, and Figure \ref{fig:features} shows the effect of the last of them pool by pool. For the v3 pools, we read the direction and the price from the signed amounts: the taker direction is the sign of the pool's signed USDC amount, $q_j = \operatorname{sign}(\text{amount0}_j)$, positive when USDC is paid in and the taker buys ETH, and the executed price is the ratio of the traded amounts. The two traded amounts have opposite signs in at least 99.89\% of swaps, and wherever consecutive swap records chain, this direction agrees with the post-swap price path $\operatorname{sign}(P_k - P_{k-1})$, where $P_k$ is the pool's marginal price after swap $k$, on 99.88-100\% of fills. Dust swaps, below \$1 or $10^{-4}$\,ETH or without an executed price (2.5-17.5\% of swaps by pool), are removed. The pool-minus-benchmark deviation is then basis-adjusted by subtracting its own three-hundred-second rolling median, which absorbs the slow USDC/USDT basis.

\begin{figure*}[t]
\includegraphics[width=\textwidth]{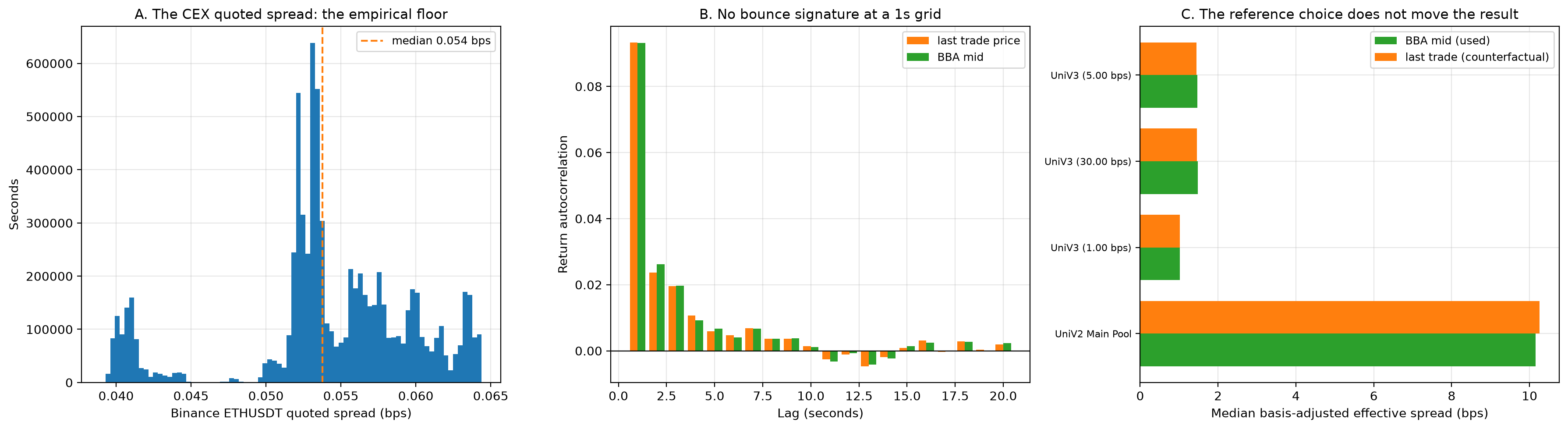}
\caption{The reference choice. (A) Distribution of the Binance quoted spread, the empirical floor, with its median marked. (B) Return autocorrelations of the last-trade series and of the best-bid-and-ask (BBA) mid: no bounce signature on a one-second grid. (C) Median basis-adjusted effective spread by pool under the mid and under the last-trade reference: the choice leaves it essentially unchanged.}
\label{fig:reference}
\end{figure*}

Panel A of Figure \ref{fig:features} shows what the basis adjustment removes: against the raw ETH-USDT mid, the median effective spread is 7.7 to 29.7\,bps across pools, and after the adjustment it is 1.0 to 10.2\,bps (panel C of Figure \ref{fig:reference}), the levels used below. The median fill pays well under twice the posted fee in every pool, as little as 2.5\% of it in the v3 30\,bps pool, because much of the flow in the high-fee pools trades when the stale price offsets the fee (see the one-sided staleness results below). Panel B is a diagnostic of the toxicity proxy we ultimately discard: the share of trades with a positive five-second markout sits at 0.34-0.40 in every pool, about half of the fills whose markout is non-zero, as expected of a statistic driven by the symmetric innovation rather than by the informed share. Finally, panel C indicates that the persistent pool-benchmark gap is a slow stablecoin basis, typically $-5$ to $-10$\, bps, that drifts over weeks, not intra-slot staleness.

\runhead{The reference choice} Since every spread and markout is measured against the benchmark mid, the choice of reference deserves scrutiny. Panel A of Figure \ref{fig:reference} shows that the Binance quoted spread is one tick wide almost always, with a median of 0.0538\, bps (and a half-spread of 0.0269\, bps) and a single tick of 0.0536 \, bps at the mean price of \$1,866.87. This is the empirical floor, i.e., the price of liquidity under no cadence constraint whatsoever, and it is the number every model spread below should be read against.

Panel B addresses the concern that a reference built from the trade tape would carry a bid-ask bounce and inflate the volatility estimate: on a one-second grid, there is no bounce to remove. The first-order return autocorrelation is $+0.093$ for both the last-trade and the mid series, and the two volatility estimates coincide to three decimals (0.851\,bps$/\sqrt{\text{s}}$). The positive autocorrelation is a departure from the i.i.d. increments of Section \ref{sec:model}, and it is not an artifact of carrying quotes forward, because the last-trade series shares it: the standard deviation of twelve-second innovations is 3.34 bps, 13\% above $\hat{\sigma} \sqrt{12} = 2.95$ bps. The empirical spreads and thresholds below use the empirical law at each horizon and so do not rely on independent increments; the Gaussian benchmark does, and with the realized twelve-second variance its low-toxicity threshold would be 74.5\% rather than 71.1\%. Panel C closes the question by re-pricing every pool against the last trade instead of the mid and finding the basis-adjusted effective spread essentially unchanged. We use the mid throughout because the effective spread is conventionally measured against the midpoint; that the choice does not move the result is itself part of the evidence.

\runhead{Identification} Every causal statement rests on one assumption: at the one-second to one-minute horizon, the benchmark mid $m_t$, its increments $Z_h$, and the drift $\hat{\mu}_t$ are predetermined with respect to any single pool's idiosyncratic order flow. The justification is a size and speed asymmetry: Binance turns over orders of magnitude more volume than the four pools and leads price discovery, so the pools are price-takers that re-converge to the benchmark through arbitrage, and benchmark variation is, from a pool's viewpoint, as good as randomly assigned. This single assumption supplies three identifying resources, each used by a different measurement: an exogenous innovation, which lets us calibrate the shock $\kappa$ without any endogenous on-chain price variation; an exogenous drift sign, which is the variation behind the directional tests; and an exogenous markout reference, which severs the mechanical link between a trade and its own price impact. A slot-level lead-lag regression confirms this ordering directly. Regressing the twelve-second return of the v3 1\,bp pool on Binance twelve-second returns from five slots before to five slots after (647,989 slots), the same-slot coefficient is 0.60, Binance returns over the preceding minute add 0.36 (standard error 0.009), and Binance returns after the pool's add only 0.007 (0.006): the pool absorbs 96\% of a Binance move within a minute and does not lead it.

\runhead{Estimation} Toxicity, the input every downstream object depends on, is disciplined by the batch-netting residual $\nu(h)$: pooling the benchmark trade tape into $h$-second batches, $\nu(h)$ is the share of volume that fails to net inside a batch, in excess of a sign-randomized null, and at short $h$, where informed flow within a batch runs one way, it bounds the informed share $\theta$ of Section \ref{sec:model} from above (parent-order splitting makes even uninformed flow one-sided there). We therefore report every $\theta$-dependent object across informed shares, $\theta \in \{0.10, 0.25, 0.50, 0.75, 0.90\}$, and mark the tightest bound, $\nu(60\,\text{s})$, and the one-second residual, $\nu(1\,\text{s})$.

Because the innovation is discrete and heavy-tailed, the truncated loss $A_h(s) = \mathbb{E}\!\left[(|Z_h|-s)_+\right]$ in the break-even condition $s^* = \theta A_h(s^*)/(1-\theta)$ is evaluated on the empirical law of $Z_h$, with the Gaussian of Proposition \ref{prop:threshold-gaussian} shown for comparison. Unless noted otherwise, standard errors are clustered by hour, since fills in the same hour share one benchmark path across pools. Where a null is needed for the netting residual, we use a randomization envelope over the trade tape; for the directional tests, we shift the drift-sign series circularly by at least one hour (500 draws), which preserves its persistence and the timing of every fill. Two placebos, a calm-window restriction and this shift test, probe the directional results.

\subsection{Results}

\begin{figure*}[t]
\includegraphics[width=\textwidth]{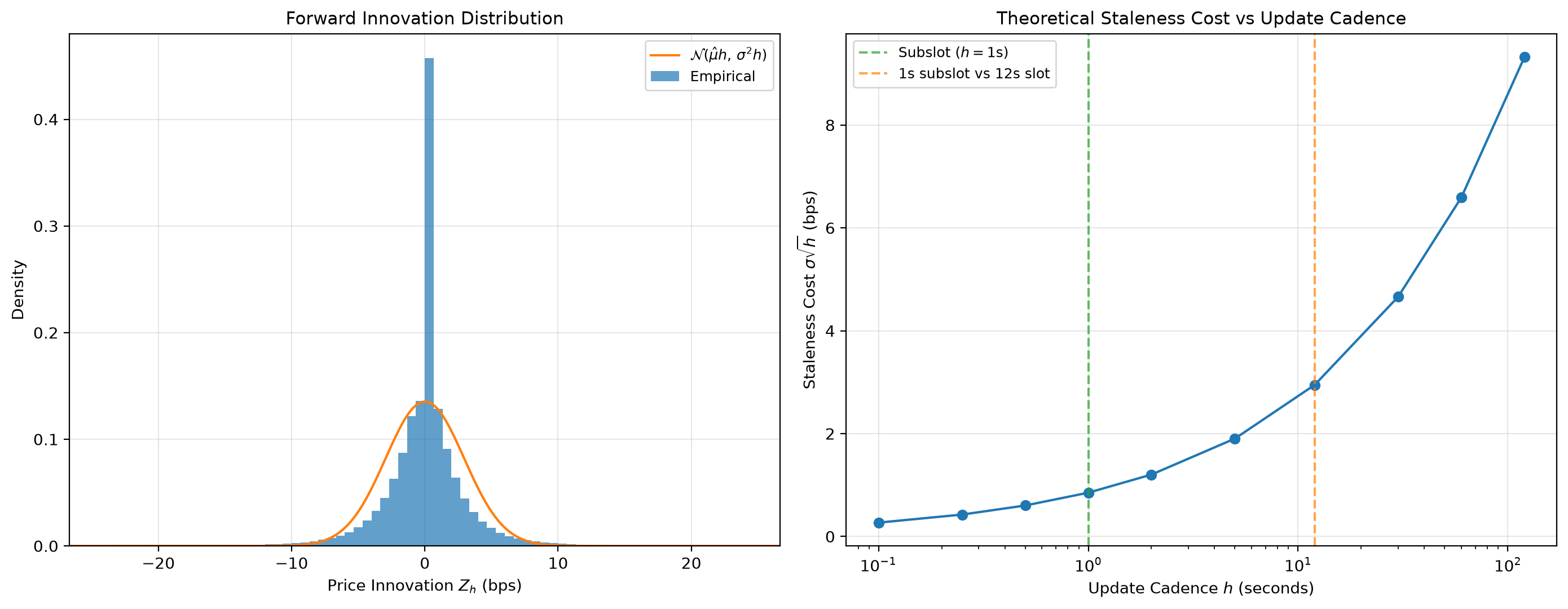}
\caption{Left: the twelve-second innovation is far more peaked than the Gaussian $\mathcal{N}(\hat\mu h, \hat\sigma^2 h)$ implied by the one-second calibration, reflecting the zero atom and fat tails. Right: the volatility anchor $\sigma \sqrt{h}$ against cadence, with the one-second subslot and the twelve-second slot marked.}
\label{fig:calibration}
\end{figure*}

\runhead{Calibration and the innovation} We estimate a one-second volatility of $\hat{\sigma} = 0.85\,\text{bps}/\sqrt{\text{s}}$ (47.8\% annualized). Conditioning on the regime, the trending volatility is $\hat{\sigma}_\text{trend} = 1.82\,\text{bps}/\sqrt{\text{s}}$ against $\hat{\sigma}_\text{range} = 0.63\,\text{bps}/\sqrt{\text{s}}$, so the trending state is $2.9\times$ as volatile and the pooled figure understates it by more than a factor of two. The subslot innovation is, however, sharply non-Gaussian in two ways that we quantify rather than set aside (Figure \ref{fig:calibration} shows both at the twelve-second horizon). First, it is heavily discrete, with a zero atom no Gaussian can place: at a one-second cadence, 73.4\% of subslots see no change in the mid, which moves only when the best bid or ask price changes, and in most seconds it does not. Seconds without a quote update, in which the last quote is carried forward, are zero changes by construction, but they are 4.5\% of the sample, and on seconds with an observed quote the atom is still 71.4\%, so they account for only 2.0 of these 73.4 percentage points. This is not a tick-coarseness effect: one tick, 0.054\, bps, is sixteen times finer than $\hat{\sigma}$, so a Gaussian of the same volatility would leave only a few percent of seconds within a tick of zero, not three-quarters. The atom reflects how rarely the mid updates; $\hat{\sigma}^2$ is carried by the seconds in which it does move, when it jumps roughly twenty ticks, so $\hat{\sigma} \sqrt{h}$ measures the tail rather than the typical second, which is why the spreads and thresholds reported below are computed on the empirical law, with the Gaussian shown for comparison. Second, it is fat-tailed, with an excess kurtosis of 154 at one second.

The joint consequence is that the Gaussian closed form overstates the typical shock at short horizons: the realized mean absolute innovation is only $\kappa(1\,\text{s}) = 0.29$\, bps, 0.43 of the Gaussian value $\hat{\sigma} \sqrt{2h/\pi}$ there, a ratio that climbs to 0.80 at twelve seconds and 0.86 at thirty as aggregation averages the calm seconds and the bursts. Drift enters the shock through the folded-normal mean of Proposition \ref{prop:drift-gaussian}. With $g(\rho):= \mathbb{E}|Z_h| / (\sigma \sqrt{h})$ and $\rho:= |\mu|\sqrt{h}/\sigma$ the per-interval signal-to-noise ratio, $g(0) = \sqrt{2/\pi}$ and a trend doubles the shock relative to a driftless interval at $\rho = 1.54$.

\runhead{The cost of staleness} Table \ref{tab:shock} prices the shock as a function of cadence under regime-conditional Gaussian calibration. Shortening cadence from twelve seconds to one lowers the expected shock by 71.1\% in both regimes: the venue that quotes fastest sheds most of the risk, in calm and trending markets alike. On the pooled empirical innovation, the reduction is larger, 84.5\% ($\kappa$ falls from 1.880 to 0.292\, bps), because $\kappa$ falls faster with cadence than $\sqrt{h}$ (see the screening threshold below); by Proposition \ref{prop:threshold-robust-comparisons}, this proportional reduction is also the low-toxicity screening threshold reported below. These are calibration results, immune to the endogeneity of on-chain trade timing. They condition on the realized regime, and the drift in Table \ref{tab:shock} is the forward drift of each regime: the benchmark move over the next twelve seconds in the direction of the trailing sixty-second return, 0.0088\,bps/s in trending and 0.0042\,bps/s in range-bound seconds. The mean absolute trailing drift is not a forward drift, because selecting seconds on a large trailing return produces it mechanically: the same one-second returns shuffled within each day, which carry no drift, give 0.250 and 0.056\,bps/s through the same cut, against the 0.284 and 0.051 measured.

\begin{table}[t]
\caption{Expected shock $\kappa(h)$ by regime (bps) and its reduction from 12\,s to 1\,s.}
\label{tab:shock}
\centering
\begin{tabular*}{\columnwidth}{@{}l@{\extracolsep{\fill}}cccc@{}}
\toprule
Regime & $\kappa(1\,\text{s})$ & $\kappa(4\,\text{s})$ & $\kappa(12\,\text{s})$  & Cut, 12\,s to 1\,s\\
\midrule
Range-bound & 0.504 & 1.009 & 1.748 & $-71.1$\% \\
Trending & 1.455 & 2.910 & 5.041 & $-71.1$\% \\
\bottomrule
\end{tabular*}
\tabnotes{$\kappa(h)=\mathbb{E}|Z_h|$ is the expected absolute innovation over $h$ seconds, from the folded-normal formula of Proposition \ref{prop:drift-gaussian} at regime-conditional estimates on the Binance ETH-USDT mid: $\hat\sigma = 0.63$\,bps$/\sqrt{\text{s}}$ and $|\hat\mu| = 0.0042$\,bps/s in the range-bound regime; $\hat\sigma = 1.82$\,bps$/\sqrt{\text{s}}$ and $|\hat\mu| = 0.0088$\,bps/s in the trending regime, where $|\hat\mu|$ is the forward benchmark move over the next twelve seconds in the direction of the trailing sixty-second return, divided by twelve. The last column is $\kappa(1\,\text{s})/\kappa(12\,\text{s})-1$; it equals the driftless Gaussian value $1/\sqrt{12}-1$ in both regimes, and the pooled empirical innovation gives $-84.5$\%.}
\end{table}

\begin{figure*}[t]
\includegraphics[width=\textwidth]{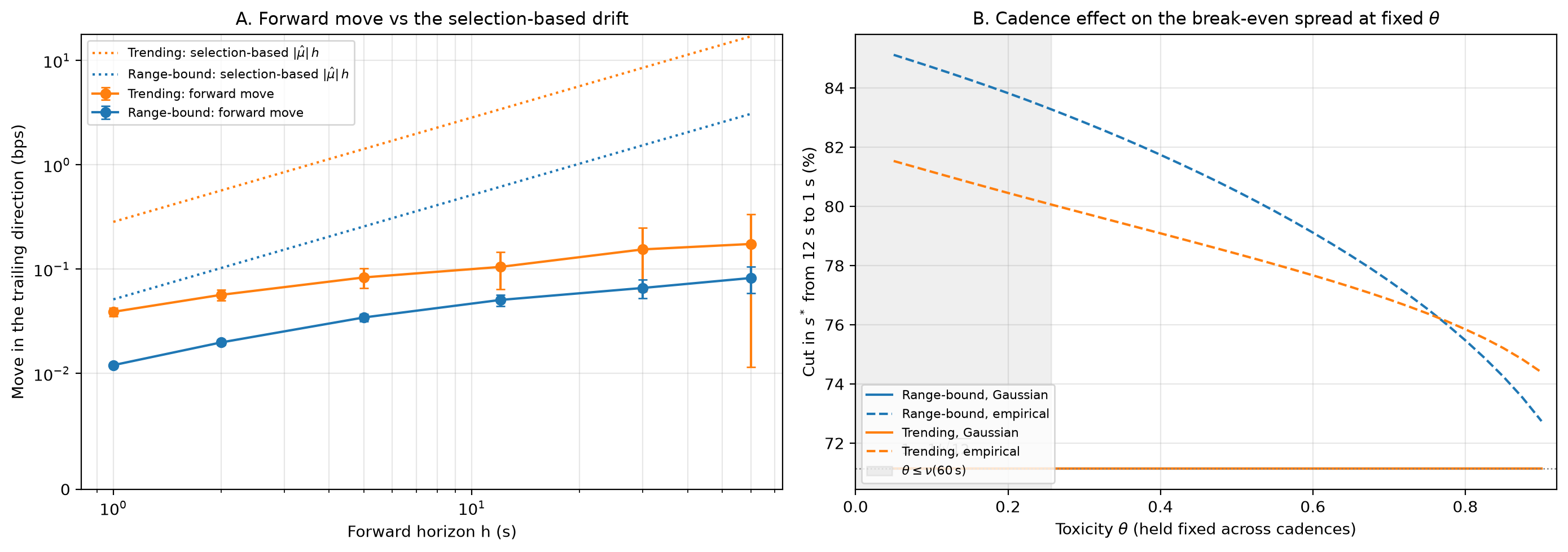}
\caption{Left: forward benchmark move in the direction of the trailing sixty-second return, by horizon and regime, with 95\% intervals from hour-clustered errors, against the move $|\hat\mu| h$ implied by the mean absolute trailing drift (dotted). Right: cut in the participation-constrained break-even half-spread $s^*$ from a twelve-second to a one-second cadence, at toxicity $\theta$ held fixed across cadences, on the empirical innovation (dashed) and under the Gaussian (solid, 71.1\% in both regimes); the shaded band is $\theta \le \nu(60\,\text{s})$.}
\label{fig:drift}
\end{figure*}

Drift is therefore a negligible share of the shock's \textit{magnitude}; its role in the model is directional. Write $\omega(h) := 1 - g(0)/g(\rho_h)$, with $\rho_h := |\mu|\sqrt{h}/\sigma$, for the drift share of the expected shock at cadence $h$. In the trending regime, drift accounts for $\omega(12\,\text{s}) = 0.01\%$ of the shock at twelve seconds and $\omega(1\,\text{s}) = 0.001\%$ at one second, because drift enters the innovation linearly in $h$ while diffusion enters as $\sqrt{h}$, as stated in Proposition \ref{prop:drift-gaussian}. The trending share reaches $1-\sqrt{2/\pi} \approx 20.2\%$, the level at which the expected shock equals the volatility anchor $\sigma\sqrt{h}$ (that is, $g(\rho_h)=1$), only at $h^* \approx 6.4$ hours. At block and subslot cadence, therefore, the risk is almost entirely diffusion. What the benchmark does carry is a short-lived continuation: in trending seconds it moves a further 0.039\,bps over the next second, 0.083\,bps over five, and 0.105\,bps over twelve in the direction of the trailing move (left panel of Figure \ref{fig:drift}), so its per-second rate falls with the horizon. The directional content that the one-sided-staleness test below detects survives because the informed trade on the side of the move that the stale quote has not yet absorbed, whatever its share of the magnitude.

\runhead{Staleness and the accumulated discrepancy} The calibration prices the shock from the benchmark alone; the pools themselves show how the discrepancy a stale quote must absorb grows with the time it has been exposed. For each block in which pool $v$ trades, with elapsed staleness $a_j$ as in \eqref{eq:staleness}, we estimate
\begin{equation}\label{reg:stale}
    \log D_j = \alpha_v + \beta \log a_j + u_j,
\end{equation}
where $D_j$ is either the absolute benchmark move since the pool's previous update or the absolute basis-adjusted deviation of the pool's pre-trade price from the benchmark (v3 pools). Under a driftless diffusion $\beta = 1/2$, rising toward one as drift comes to dominate. Elapsed staleness is endogenous, because arbitrageurs update a pool sooner after a large benchmark move, which biases OLS downward. We therefore instrument $\log a_j$ with the log block gap: a missed proposer slot lengthens the gap from twelve seconds to twenty-four or thirty-six for reasons unrelated to the market. Table \ref{tab:stale} reports the estimates. The first stage is strong ($F = 3{,}046$), and the instrumented elasticity of the benchmark move is 0.45 (standard error 0.05), in line with the diffusion rate of one-half and with the benchmark's own scaling of 0.56 between twelve and sixty seconds; its point estimate is higher in trending markets (0.60) than in range-bound ones (0.41), the ordering the model predicts. The pool's own pre-trade deviation grows with an elasticity of 0.32 (0.04). OLS lies below IV in every column, as the endogeneity of update timing implies. Each additional second of staleness therefore raises the discrepancy a pool must absorb at close to the diffusion rate, which is the cost that subslot cadence removes.

\begin{table}[t]
\caption{Staleness elasticity of the stale-quote discrepancy.}
\label{tab:stale}
\centering
\begin{tabular*}{\columnwidth}{@{}l@{\extracolsep{\fill}}ccc@{}}
\toprule
& All & Range-bound & Trending \\
\midrule
\multicolumn{4}{@{}l}{\textit{Benchmark move since the previous update}} \\
IV & $0.448^{***}$ & $0.408^{***}$ & $0.595^{***}$ \\
 & $(0.054)$ & $(0.055)$ & $(0.099)$ \\
OLS & $0.219^{***}$ & $0.288^{***}$ & $0.437^{***}$ \\
 & $(0.005)$ & $(0.004)$ & $(0.007)$ \\
First-stage $F$ & 3,046 & 2,506 & 658 \\
Observations & 791,264 & 660,426 & 130,838 \\
\midrule
\multicolumn{4}{@{}l}{\textit{Pre-trade pool deviation (v3 pools)}} \\
IV & $0.318^{***}$ & $0.273^{***}$ & $0.412^{***}$ \\
 & $(0.043)$ & $(0.044)$ & $(0.072)$ \\
OLS & $-0.059^{***}$ & $-0.026^{***}$ & $0.148^{***}$ \\
 & $(0.003)$ & $(0.003)$ & $(0.005)$ \\
First-stage $F$ & 3,900 & 3,357 & 713 \\
Observations & 740,866 & 624,784 & 116,082 \\
\bottomrule
\end{tabular*}
\tabnotes{Regressions \eqref{reg:stale} of the log discrepancy on log elapsed staleness with pool fixed effects, over blocks in which the pool trades and has a previous update. IV instruments $\log a_j$ with the log block gap, the time since the previous block (12, 24, or 36 seconds); the gap exceeds twelve seconds after a missed proposer slot, in 0.38\% of pool-blocks. OLS and IV use pool-blocks with a positive discrepancy (87.5\% of the benchmark-move sample and all of the pool-deviation sample). Standard errors clustered by hour are in parentheses. $^{*}p<0.10$, $^{**}p<0.05$, $^{***}p<0.01$.}
\end{table}

\runhead{The endogenous spread} The half-spread $s^*$ of Proposition \ref{prop:spread-existence} is participation-constrained, and it departs from the volatility-anchored rule of thumb $\theta\sigma\sqrt{h}/(1-\theta)$, which charges as if every informed order traded against a shock of size $\sigma\sqrt{h}$ and so ignores that an informed order executes only when $|Z_h| > s$. Netting out the deterred orders replaces $\sigma\sqrt{h}$ with the truncated loss $A_h(s^*)$, and the rule of thumb overstates $s^*$ by a factor that grows with toxicity, from about 1.7$\times$ at $\theta=0.10$ to about 7$\times$ at $\theta=0.90$ on the empirical law (Table \ref{tab:spread}). Part of the gap at low toxicity comes from the anchor itself rather than from participation: $\sigma\sqrt{h}$ exceeds $\mathbb{E}|Z_h|$ by the factor $\sqrt{\pi/2}\approx1.25$ under the Gaussian, and by more on the empirical law.

\begin{table}[t]
\caption{Break-even half-spread at $h = 12$\,s (bps): naive rule vs.\ participation-constrained, Gaussian vs.\ empirical.}
\label{tab:spread}
\centering
\begin{tabular*}{\columnwidth}{@{}c@{\extracolsep{\fill}}ccc@{}}
\toprule
$\theta$ & Naive $\frac{\theta}{1-\theta}\sigma\sqrt{h}$ & Gaussian & Empirical \\
\midrule
0.10 & 0.33 & 0.24 & 0.19 \\
0.25 & 0.98 & 0.60 & 0.51 \\
0.50 & 2.95 & 1.29 & 1.15 \\
0.75 & 8.84 & 2.26 & 2.24 \\
0.90 & 26.53 & 3.36 & 3.82 \\
\bottomrule
\end{tabular*}
\tabnotes{Half-spreads without screening ($\eta = 0$). The naive column uses $\hat\sigma\sqrt{12} = 2.95$\,bps. The Gaussian and empirical columns solve the break-even condition~\eqref{eq:spread-break-even}: the former on the driftless closed form of Proposition \ref{prop:threshold-gaussian} with $\hat\sigma = 0.85$\,bps$/\sqrt{\text{s}}$, the latter on 647,999 non-overlapping twelve-second innovations of the Binance ETH-USDT mid. The two cross at $\theta \approx 0.76$.}
\end{table}

Evaluating $A_h$ on the empirical law rather than the Gaussian is not a one-way correction: the two cross just above $\theta = 0.75$; the empirical spread is tighter below (the zero atom thins the body) and wider above (the heavier tail feeds the region $|Z_h| > s$). Holding toxicity fixed across cadences isolates the effect of cadence (right panel of Figure \ref{fig:drift}): going from twelve seconds to one cuts the empirical break-even half-spread by 85\% at $\theta = 0.10$ and by 73\% at $\theta = 0.90$ in the range-bound regime, and by 81\% to 74\% in the trending regime, against 71\% under the Gaussian. At $\theta = 0.50$, for instance, the empirical half-spread falls from 0.922 to 0.180\,bps range-bound and from 2.747 to 0.593\,bps trending. At $\theta = \nu(1\,\text{s})$, the unscreened one-second half-spread, 0.454\,bps, remains more than ten times the 0.027\,bps half-spread quoted on the CEX. Drift raises this spread only at second order, as stated in Proposition \ref{prop:drift-local}: a small predictable component has no first-order effect on the symmetric, balanced-flow spread.

\runhead{Toxicity} The batch-netting residual falls monotonically with batch length, from $\nu (1\,\text{s}) = 0.706$ through $\nu(12\,\text{s}) = 0.402$ to $\nu(60\,\text{s}) = 0.257$, as longer batches internalize more offsetting flow; at short horizons it is an upper bound on the informed share, since parent-order splitting makes even benign flow arrive one-sided there. How far $\nu$ can be pushed as a toxicity estimate is itself bounded by the null: at twelve seconds, the raw imbalance of 0.509 exceeds a global sign shuffle by 0.330, a shuffle of signs within five-second windows by 0.049, and a cyclic rotation of the sign sequence, which preserves runs and hence autocorrelation and order splitting, by $-0.007$, i.e., by nothing. A run-preserving null therefore explains almost all of the directional imbalance, so $\nu$ bounds toxicity from above rather than identifying it. We therefore treat $\theta$ as a parameter, with $\nu(60\,\text{s}) = 0.257$ as the binding upper bound, and, as the screening threshold below shows, the conclusions hold across this range. Appendix \ref{app:extended} sets each pool's posted fee against its post-trade markout and against the revenue its providers keep at execution.

\begin{figure*}[t]
\includegraphics[width=\textwidth]{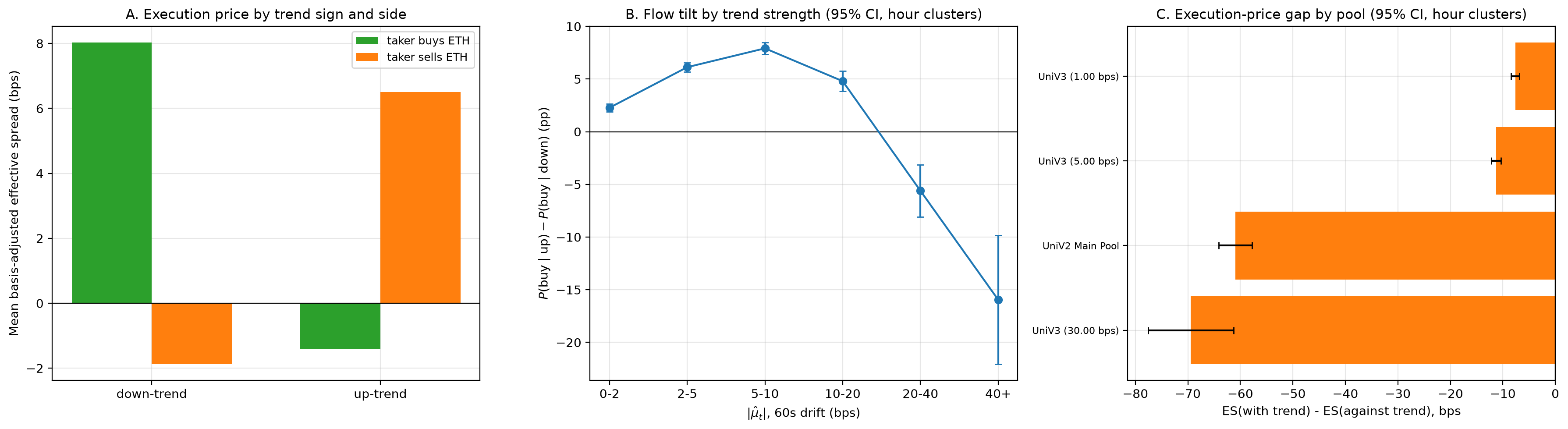}
\caption{One-sided staleness in trending markets. (A) Mean basis-adjusted effective spread by trend sign and trade side: with-trend fills execute below the benchmark mid. (B) Share of taker buys in up-trends minus down-trends across bins of the absolute sixty-second drift, with 95\% intervals from hour-clustered errors: the with-trend tilt widens up to 10\, bps, narrows, and reverses beyond 20\, bps. (C) Effective spread of with-trend minus against-trend fills by pool, with 95\% intervals.}
\label{fig:markout}
\end{figure*}

\runhead{One-sided staleness in trending markets} The model's sharpest prediction concerns the direction of the staleness cost. The informed choose $D_i = \operatorname{sign}(Z_h(\mu))$ and execute when $|Z_h(\mu)|$ is large, so in a trend they hit the stale quote on the side toward which the price is moving: executed informed orders are disproportionately with-trend, and the with-trend side carries the larger expected loss. Panel A of Figure \ref{fig:markout} confirms the loss asymmetry in the data and shows where it is paid: with swaps timed at their block, the stale quote is picked off in the execution price itself.

The pooled pick-off gap, i.e., the difference in effective spread between with-trend and against-trend fills in the trending regime, equals $-11.1$\,bps with pool fixed effects (hour-clustered standard error 0.41, $t = -27.3$; shift test $p < 0.002$) and $-8.68$ bps in the raw means: with-trend fills execute at $-1.71$\,bps, better than the benchmark mid (median $-1.92$), while against-trend fills pay $6.97$\,bps (median 3.67). The gap has the predicted sign and is significant in all four pools: $-7.6$\,bps in the v3 1\,bp pool, $-11.3$\,bps in the v3 5\,bps pool, $-69.4$\,bps in the v3 30\,bps pool, and $-60.9$\,bps in the v2 pool. In the two thirty-basis-point pools it is about twice the fee, because with-trend flow there trades when the stale price covers the fee, while against-trend flow pays the fee in full. With-trend takers also arrive after a larger benchmark move: over the twelve seconds before the block, the benchmark moves 4.85\,bps in the trend's direction before with-trend fills and 2.77\,bps before against-trend fills (a raw difference of 2.08\,bps and 1.77\,bps with pool fixed effects, standard error 0.09). Once the block lands, the benchmark does not move further in the taker's favor: the five-second markout gap is $-0.06$\,bps and not significant, so the stale-quote loss is realized at execution rather than after it.

We put the pick-off and the flow tilt on a regression footing (Table \ref{tab:tab4}). The pick-off regressions relate a fill-level outcome to the drift-aligned trade sign on trending fills, with pool fixed effects and hour clusters:
\begin{equation}\label{reg:reg3}
    Y_j = \alpha_v + \lambda w_j + u_j, \qquad w_j = q_j \operatorname{sign}(\hat{\mu}_{t_j}),
\end{equation}
where $Y_j$ is the effective spread $\mathrm{ES}_j$, the benchmark move over the slot before the block, $\mathrm{PM}_j = q_j (m_{t_j} - m_{t_j - 12})/m_{t_j - 12}$, or the markout $\mathrm{MO}_j(5\,\text{s})$, so that $2\lambda$ is the with-trend minus against-trend gap. The flow tilt is a clustered logit of the buy indicator on the drift:
\begin{equation}\label{reg:reg2}
    \mathbb{P} (q_j = 1) = \Lambda (\alpha_v + \beta \hat{\mu}_{t_j}),
\end{equation}
where $\Lambda$ is the logistic function.

\begin{table}[t]
\caption{Directional regressions (fill level).}
\label{tab:tab4}
\centering
\begin{tabular*}{\columnwidth}{@{}l@{\extracolsep{\fill}}cccc@{}}
\toprule
& (1) & (2) & (3) & (4) \\
& $\text{ES}$ & $\text{PM}$ & $\text{MO}(5\,\text{s})$ & Buy \\
\midrule
$w=q\,\operatorname{sign}(\hat\mu)$ & $-5.542^{***}$ & $3.594^{***}$ & $-0.029$ & \\
 & $(0.203)$ & $(0.063)$ & $(0.028)$ & \\
$\hat\mu$ & & & & $0.0005$ \\
 & & & & $(0.0006)$ \\
\midrule
Gap $2\lambda$ (bps) & $-11.09$ & $7.19$ & $-0.06$ & \\
Estimator & OLS & OLS & OLS & Logit \\
Sample & Trending & Trending & Trending & All \\
Pool FE & Yes & Yes & Yes & Yes \\
Cluster & hour & hour & hour & hour \\
Observations & 318,822 & 318,822 & 318,822 & 1,725,527 \\
(Pseudo-)$R^2$ & $0.016$ & $0.199$ & $0.0001$ & $0.001$ \\
\bottomrule
\end{tabular*}
\tabnotes{Columns (1)-(3) are OLS regressions \eqref{reg:reg3} of a fill-level outcome on $w=q\,\operatorname{sign}(\hat\mu)$ over trending fills, so that twice the coefficient is the with-trend minus against-trend gap; column (4) is a logit \eqref{reg:reg2} over all fills with a trailing sixty-second return (twelve of the 1,725,539 have none) whose coefficient is in log-odds (average marginal effect on $\mathbb{P}(\text{buy})$: $+0.00013$ per bp of drift). ES is the basis-adjusted effective spread, PM the benchmark move over the twelve seconds before the block signed by the taker's side, and MO(5\,s) the five-second markout; all are in bps, as is $\hat\mu$. All columns include pool fixed effects; standard errors clustered by hour are in parentheses. $^{*}p<0.10$, $^{**}p<0.05$, $^{***}p<0.01$.}
\end{table}

The effective-spread coefficient carries the pick-off: with-trend fills pay 11.1\,bps less than against-trend fills (column (1)), and the gap is unchanged with day clusters (standard error 0.41) and survives the shift test (below). Column (2) shows the selection behind it: with-trend takers act after the benchmark has moved their way. The post-trade markout in column (3) is flat, as expected once execution is timed at the block: the benchmark move that makes the quote stale precedes the block and does not continue in the taker's direction after it. The flow tilt follows the trend at moderate drifts. In the trending regime, takers are 0.9 percentage points more likely to buy in an up-trend than in a down-trend ($\mathbb{P}(\text{buy}\mid\text{up}) = 0.328$ vs.\ $\mathbb{P}(\text{buy}\mid\text{down}) = 0.318$). Across all fills, by drift bin, the tilt rises to 7.9 points (standard error 0.3) at 5-10\,bps, below the trending cut, before reversing once the sixty-second drift exceeds 20\, bps ($-5.6$ points at 20-40\, bps, $-15.9$ points above 40\, bps; panel B of Figure \ref{fig:markout}). Because a linear logit averages these opposite responses, column (4) is not significant, and the directional claim rests on the loss asymmetry, measured in the execution price, not on the tilt.

\begin{figure}[t]
\includegraphics[width=\columnwidth]{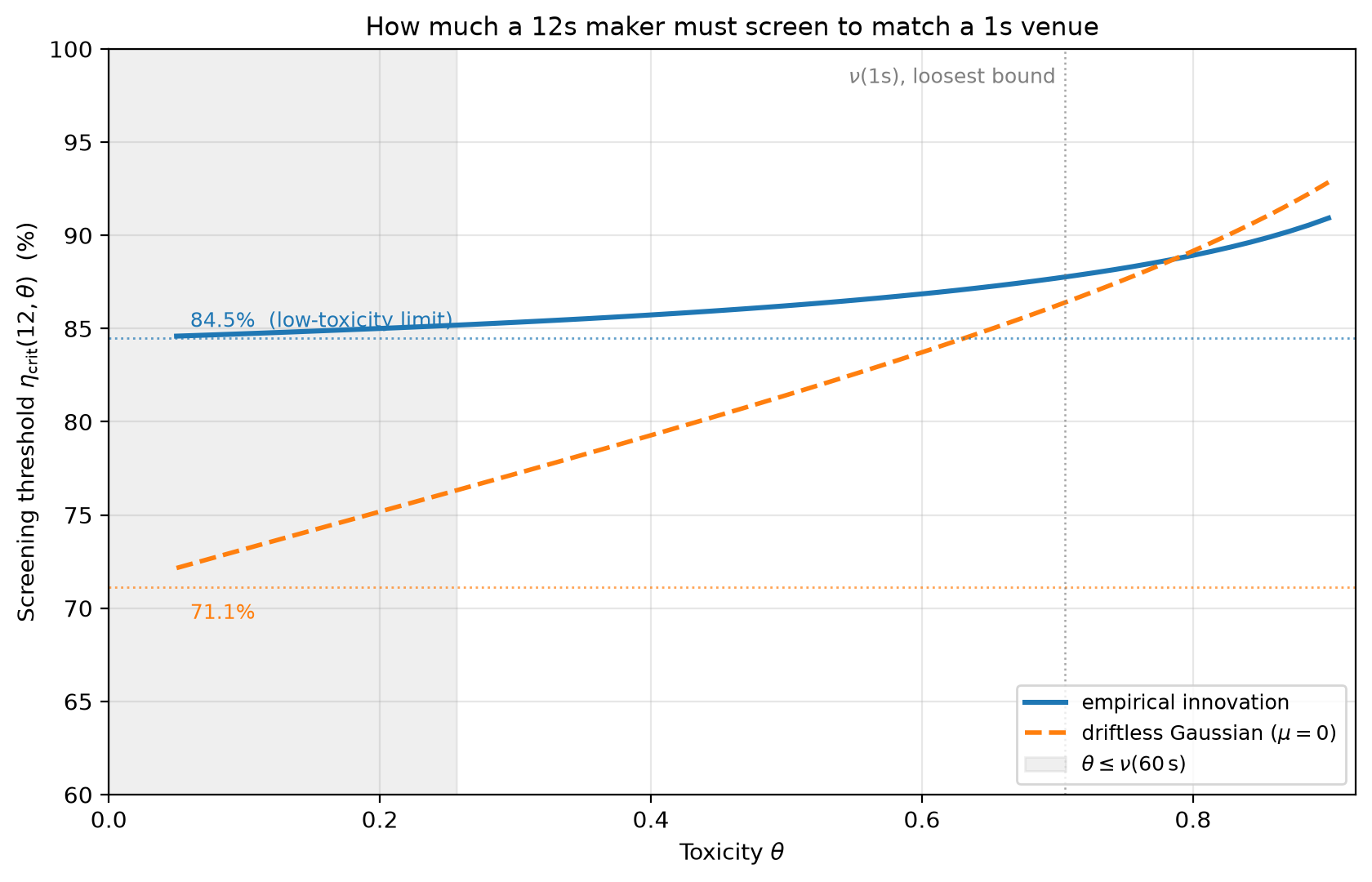}
\caption{The screening threshold $\eta_{\mathrm{crit}}(12,\theta)$: how much a twelve-second maker must screen to match a one-second venue, on the empirical innovation versus the driftless Gaussian. Both rise with toxicity. The empirical curve starts 13 percentage points higher at the low-toxicity limit and is the flatter of the two, so the curves nearly meet and cross at high toxicity. The shaded band is $\theta \le \nu(60\,\text{s}) = 0.257$, the tightest netting bound on toxicity; the dotted line marks $\nu(1\,\text{s}) = 0.706$.}
\label{fig:comparison}
\end{figure}

\runhead{The screening threshold} The model's central comparison, Theorem \ref{prop:screening-threshold}, has a direct empirical counterpart: the rate at which a twelve-second maker must screen to match a one-second venue (Figure \ref{fig:comparison}). In the low-toxicity limit this is $\eta_{\mathrm{crit}} = 1 - \mathbb{E}|Z_{h_0}|/\mathbb{E}|Z_{h_P}| = 1 - \kappa(1\,\text{s})/\kappa(12\,\text{s})$. Evaluated on the empirical innovation, it equals $1 - 0.292/1.880 \approx 84.5\%$, 13.3 percentage points above the driftless Gaussian benchmark of $1 - 1/\sqrt{12} \approx 71.1\%$, because $\kappa$ falls faster with cadence than $\sqrt{h}$ would predict. Positive serial dependence at the one-second horizon accounts for about 3 points of the gap ($1 - 1/\sqrt{12 \times 1.285} = 74.5\%$, with 1.285 the twelve-second variance ratio), and the shape of the one-second innovation for the remaining 10: most seconds are calm and a minority carry large moves, which aggregation to twelve seconds averages out. The threshold does not hinge on how the reference is built: across the BBA mid, the mid on seconds with an observed quote, the microprice (whose zero atom is only 5\%), a trade-implied mid, and the last trade, it ranges from 84.0\% to 85.1\%. Solving the exact break-even at positive toxicity raises both curves and, because the empirical curve is the flatter of the two, they nearly meet at high toxicity: at $\theta = \nu(1\,\text{s}) = 0.706$ the empirical threshold is 87.8\% against the Gaussian 86.4\%, and the two cross near $\theta \approx 0.78$. Over the informed shares admitted by the tightest netting bound, $\theta \le \nu(60\,\text{s}) = 0.257$, the empirical threshold stays between 84.5\% and 85.2\%, and from the low-toxicity limit up to $\theta = 0.9$ between 84.5\% and 90.9\%, so the conclusion does not depend on pinning toxicity down.

Either way, the reading is the same: screening would have to be extraordinarily effective, rejecting on the order of 85\% of informed opportunities, and more at higher toxicity, for a block-cadence maker to match one-second execution on this pair. The equal-cadence premium quantifies the converse: at $\theta = \nu(1\,\text{s})$ a one-second maker that screens even a quarter of informed flow already prices 0.084\, bps inside an unscreened one-second venue, whose half-spread is 0.454\, bps, rising to 0.302\, bps at three-quarters screening.

\runhead{Robustness} Four checks discipline the directional results. The shift test is the sharpest and most assumption-light of them. Moving the drift-sign series circularly by at least an hour preserves its persistence and the timing of every fill while breaking its alignment with the fills; over 500 such shifts, the effective-spread gap has a null centered at zero (mean $+0.005$\,bps, standard deviation 0.30\,bps), and no draw exceeds 1.1\,bps in absolute value against the observed $-11.1$\,bps ($p < 0.002$).

The forward-drift check asks how much of the drift is known in advance, which is what the recentering result of Proposition \ref{prop:drift-recentering} requires. On a chronological holdout (the last 40\% of the sample, 3.1 million seconds), the benchmark return in the direction of the trailing sixty-second drift averages $+0.014$\, bps over the next second and $+0.064$\, bps over the next twelve, small although statistically different from zero. Unconditionally, then, $\hat{\mu}_t$ describes a move that has largely already occurred rather than one about to occur. Within the trending seconds of the holdout, the continuation is $+0.040$, $+0.095$, and $+0.130$\,bps over one, five, and twelve seconds, 14\%, 7\%, and 4\% of the move that the mean absolute trailing drift would imply (0.28, 1.4, and 3.4\,bps). Predictable drift is therefore small in this sample, so the pick-off documented above reflects staleness rather than drift, and a recentered quote would gain little over a spot-centered one.

The calm-window placebo is passed: restricted to $|\hat{\mu}_t| < 2$\, bps, the five-second markout gap is 0.041\,bps (standard error 0.011), matching the 0.038\,bps that the benchmark's own continuation over the same seconds produces, so outside trends the markout gap is accounted for by continuation. The slot-level lead-lag test confirms the identifying assumption: the pool absorbs 60\% of a Binance move in the same slot and 96\% within a minute, and its own returns do not lead Binance (0.007, standard error 0.006).

\section{Conclusion}\label{sec:conclusion}

This paper studies the spread a liquidity provider requires to break even against orders that can exploit an outdated quote. That spread depends on two features of the trading arrangement: the time for which the quote remains binding as the reference price moves and the maker's ability to reject informed orders, and we place these two protections on a single scale. The competitive half-spread is the unique spread at which revenue from uninformed flow covers the expected excess of the stale-price innovation over the spread, weighted by the ratio of admitted informed to uninformed flow. Shortening the horizon lowers that expected excess at every spread, screening lowers the ratio, and Theorem \ref{prop:screening-threshold} gives the exact screening rate at which a slower maker matches a faster unscreened venue: the share of the slower maker's expected loss, evaluated at the faster venue's spread, that must be removed. The comparison requires only an integrable innovation, so it holds with jumps, atoms, and infinite variance.

Three theoretical results shape how the threshold should be read. First, it is scale-free and converges, as informed flow becomes rare, to one minus the ratio of expected absolute innovations at the two horizons, which under Gaussian innovations is one minus the square root of the horizon ratio. Second, its response to toxicity is governed by the expected loss per executed informed order rather than by aggregate loss. Convexity orders aggregate losses across horizons but not this conditional severity, so in general the threshold can fall as toxicity rises. The Gaussian case is what delivers a threshold that increases with toxicity, together with lower executed volume for the screened maker at the matching spread. Third, at equal horizons, screening narrows the spread by strictly less than the screening rate times the unscreened spread, because the narrower spread makes additional informed opportunities profitable.

Predictable drift is a centering problem rather than a spread problem. When the quote is held at the current reference, drift raises the break-even spread, at second order near zero drift under Gaussian innovations and linearly in the exposure horizon when residual uncertainty vanishes. In the Gaussian low-toxicity limit, it raises the required screening from one minus the square root of the horizon ratio toward one minus the horizon ratio itself as drift comes to dominate diffusion. A maker that centers its quote on the expected terminal reference removes this premium entirely and recovers the driftless spread (Proposition \ref{prop:drift-recentering}). The drift premium therefore measures the cost of a centering restriction. The symmetry of residual innovations is a substantive assumption, since without it drift can reverse the ordering of losses across horizons.

Taken to USDC-WETH execution from June to August 2026, benchmarked against the Binance ETH-USDT midpoint, the model gives a direct answer to the question in the title. The one-second innovation is far from Gaussian, with 73.4\% of seconds showing no price change and an excess kurtosis of 154, so the Gaussian closed form overstates the typical shock at short horizons, and the headline spreads and thresholds are computed on the empirical law. Under the regime-conditional Gaussian calibration, cutting cadence from twelve seconds to one lowers the expected shock by 71.1\% in both range-bound and trending markets, and measured forward drift is too small to matter at any cadence; on the empirical innovation, the reduction is 84.5\%, and at a fixed informed share the break-even half-spread falls by 73\% to 85\%. Accordingly, a twelve-second maker must reject about 84.5\% of informed opportunities to match a one-second venue in the low-toxicity limit, against 71.1\% under the Gaussian benchmark, between 84.5\% and 85.2\% for any informed share up to the sixty-second netting bound, and about 87.8\% at $\theta = 0.71$, the one-second netting residual. Matching one-second execution from a twelve-second slot is therefore worth screening out roughly six in seven informed orders. Conversely, at an equal one-second cadence, a maker that screens a quarter of informed flow prices 0.084\, bps inside an unscreened venue, and 0.302\, bps at three-quarters.

The direction of the loss also matches the model, and the stale quote is picked off at execution. In trending markets, with-trend fills execute at an effective spread 11.1\,bps below that of against-trend fills, with the same sign in all four pools, and they follow a larger benchmark move in the trend's direction over the preceding slot; once the block lands, the benchmark does not move further in the taker's favor, and the five-second markout gap is indistinguishable from zero. The v3 1\,bp pool absorbs 60\% of a Binance move within the same slot and does not lead it, and missed proposer slots show the discrepancy a stale pool must absorb growing with elapsed staleness at an elasticity of 0.45, close to the diffusion rate of one-half. Predictable drift is small in this sample, so there is little for a recentered quote to remove.

Several limitations bound these conclusions. Toxicity is not point-identified. The batch-netting residual is an upper bound whose excess over a run-preserving null is indistinguishable from zero, so we report the threshold across informed shares up to that bound, over which it moves by less than one percentage point. The model holds potential flow fixed and resets each quote to the reference, which abstracts both from the change in counterparty composition that faster repricing can induce and from the fee-determined no-arbitrage band of an AMM. Screening is costless and independent of opportunity size, whereas a classifier that preferentially rejects the most profitable informed orders would call for a loss-weighted threshold.

\bibliography{references}

\onecolumngrid
\newpage
\appendix

\section{Proofs for the zero-drift model}
\label{app:model-proofs}

\begin{lemma}[Integrated tails and the break-even root]
\label{lem:stop-loss-root}
Let $X\geq0$ satisfy $0<\mathbb E [X]<\infty$, and define
$A(s)=\mathbb E[(X-s)_+]$ for $s\geq0$. Then
\[
    A(s)=\int_s^\infty\mathbb P(X>u)\,du
\]
The function $A$ is continuous, nonincreasing, convex, and
$1$-Lipschitz, with $A(0)>0$ and $A(s)\to0$ as $s\to\infty$.
For $s>0$, its one-sided derivatives are
\[
    A'_+(s)=-\mathbb P(X>s),\quad
    A'_-(s)=-\mathbb P(X\geq s)
\]
For every $r\geq0$, $s=rA(s)$ has a unique nonnegative root.
It is zero at $r=0$, positive at $r>0$, continuous in $r$,
and strictly increasing in $r$.
If two such loss functions satisfy $A_2(s)\geq A_1(s)$ for
every $s$, their roots are ordered in the same way.
For $r>0$, the ordering is strict if
$A_2(s_1)>A_1(s_1)$ at the first root.
\end{lemma}
\begin{proof}
Tonelli's theorem applied to
$(X-s)_+=\int_s^\infty\mathbf1_{\{X>u\}}\,du$
gives the integrated-tail identity. The other properties follow
from the corresponding pointwise properties of $(X-s)_+$;
dominated convergence with dominating variable $X$ gives the
limit at infinity. Right and left difference quotients are
bounded by one, so dominated convergence gives the stated
one-sided derivatives.

For $r>0$, the function $g_r(s)=s-rA(s)$ is continuous,
strictly increasing, negative at zero, and tends to infinity.
This proves existence and uniqueness, including positivity.
For $r=0$, the root is zero. At a root with $r>0$,
$A(s)=s/r>0$. Increasing $r$ makes $g_r$ strictly negative
at the old root, proving strict monotonicity.
Continuity holds because, on any bounded set of $r$,
roots satisfy $0\leq s\leq rA(0)$; every convergent subsequence
as $r_n\to r$ solves the limiting equation, so uniqueness
identifies its limit. This also covers $r=0$.
Finally, evaluation of $s-rA_2(s)$ at the root for $A_1$
proves both loss-ordering assertions.
\end{proof}

\begin{lemma}[A density-ratio criterion for informed loss severity]
\label{lem:tail-severity-order}
Let $X_1,X_2$ have positive continuous densities $f_1,f_2$
on $[0,\infty)$ and finite means. If $f_2(x)/f_1(x)$ is
strictly increasing, then for every $s\geq0$,
\[
    \frac{\mathbb E[(X_2-s)_+]}{\mathbb P(X_2>s)}
       >
    \frac{\mathbb E[(X_1-s)_+]}{\mathbb P(X_1>s)}
\]
\end{lemma}
\begin{proof}
Write $S_j(s)=\mathbb P(X_j>s)$ and $R(s)=S_2(s)/S_1(s)$.
The ratio $R(s)$ is the conditional expectation of
$f_2(X_1)/f_1(X_1)$ given $X_1>s$. Positivity of the density
and strict increase of its ratio imply
$R(s)>f_2(s)/f_1(s)$. Therefore
\[
    R'(s)=\frac{f_1(s)}{S_1(s)}
                 \left[R(s)-\frac{f_2(s)}{f_1(s)}\right]>0
\]
For every $u>0$, this yields
$S_2(s+u)/S_2(s)>S_1(s+u)/S_1(s)$.
Integration over $u>0$, using Lemma \ref{lem:stop-loss-root},
proves the assertion. The integrals are finite by integrability.
\end{proof}

\begin{proof}[Proof of Proposition \ref{prop:spread-existence}]
The sum $Z_h$ is integrable. Conditional Jensen's inequality gives $\mathbb E|Z_h|\geq\mathbb E|\varepsilon_t|>0$: condition on the first increment and use the zero mean of the remaining independent increments. Apply Lemma \ref{lem:stop-loss-root} with $X=|Z_h|$ and
$r=\theta(1-\eta)/(1-\theta)$.
\end{proof}

\begin{proof}[Proof of Proposition \ref{prop:spread-toxicity}]
For $\eta<1$, the ratio $r(\theta,\eta)$ is strictly increasing
in $\theta$. The assertion follows from the strict monotonicity of
the root in Lemma \ref{lem:stop-loss-root}.
At $\eta=1$, the ratio is zero for every $\theta$.
\end{proof}

\begin{proof}[Proof of Proposition \ref{prop:spread-screening}]
For fixed $\theta\in(0,1)$, $r(\theta,\eta)$ is strictly decreasing
in $\eta$. Apply the same lemma, including its root at $r=0$.
\end{proof}

\begin{proof}[Proof of Proposition \ref{prop:spread-cadence}]
For fixed $s\geq0$, let $\ell_s(x)=(|x|-s)_+$.
This function is convex. Since
$Z_{h+1}=Z_h+\varepsilon_{t+h}$ with an independent mean-zero
increment, conditional Jensen's inequality yields
\[
    \mathbb E[\ell_s(Z_h+\varepsilon_{t+h})\mid Z_h]
       \geq \ell_s(Z_h)
\]
Hence $A_{h+1}(s)\geq A_h(s)$ for every $s$.
Lemma \ref{lem:stop-loss-root} gives the spread comparison.

If $\sigma>0$, each new increment has a positive density on
the real line, being a convolution with a nondegenerate Gaussian.
Conditional on any $Z_h$, its support crosses intervals on which
$\ell_s$ has different slopes. Jensen's inequality is therefore
strict. Thus $A_{h+1}(s)>A_h(s)$ for finite $s$, and the spread
comparison is strict for $r>0$, equivalently $\eta<1$.
\end{proof}

\begin{proof}[Proof of Proposition \ref{prop:threshold-robust-comparisons}]
At the benchmark root,
$A_{h_0}(s_0)=(1-\theta)s_0/\theta>0$.
Its value is independent of $h_P$, while $A_{h_P}(s_0)$ is
nondecreasing in $h_P$, strictly so in the specified diffusion case.
The threshold ratio proves the horizon assertions.

If every innovation is multiplied by $c>0$, the new loss
at spread $cs$ is $cA_h(s)$. Uniqueness implies that the new
root is $cs^*(h,\theta,\eta)$. Both terms in the threshold
ratio are multiplied by $c$, proving invariance.

Continuity of $s_0$ in $\theta$ follows from
Lemma \ref{lem:stop-loss-root} and continuity of $r(\theta,0)$.
The threshold is consequently continuous, since its denominator
is positive. Also
$0\leq s_0\leq\theta A_{h_0}(0)/(1-\theta)$, so $s_0\to0$
as $\theta\downarrow0$. Continuity of the loss functions gives
\[
    1-\frac{A_{h_0}(s_0)}{A_{h_P}(s_0)}
       \longrightarrow
    1-\frac{\mathbb E|Z_{h_0}|}{\mathbb E|Z_{h_P}|}
\]
\end{proof}

\begin{proof}[Proof of Proposition \ref{prop:threshold-toxicity-volume}]
At an argument without an atom, Lemma \ref{lem:stop-loss-root}
gives $A_h'(s)=-\mathbb P(|Z_h|>s)$. Differentiation of
$(1-\theta)s_0=\theta A_{h_0}(s_0)$ gives
\[
    \frac{ds_0}{d\theta}
      =\frac{s_0+A_{h_0}(s_0)}
             {1-\theta+\theta\mathbb P(|Z_{h_0}|>s_0)}
      =\frac{s_0}
             {\theta[1-\theta+\theta\mathbb P(|Z_{h_0}|>s_0)]}
\]
For rigor, no differentiability at neighboring arguments is needed.
Locally the profit function increases in $s$ at a rate bounded
below by a positive constant. Changing $\theta$ changes profit
locally by at most a constant times $|\Delta\theta|$, so its root
changes by $O(|\Delta\theta|)$. First-order expansion of the
equilibrium equation at the specified non-atom gives the displayed
derivative. Differentiating the threshold ratio then proves
\eqref{eq:threshold-toxicity-derivative}.

At the threshold, substitute
$1-\eta_{\mathrm{crit}}=A_{h_0}(s_0)/A_{h_P}(s_0)$ into $V$.
The difference in executed volumes is
\[
    \theta\left[
      \frac{A_{h_0}(s_0)}{A_{h_P}(s_0)}
                      \mathbb P(|Z_{h_P}|>s_0)
        -\mathbb P(|Z_{h_0}|>s_0)\right],
\]
which proves \eqref{eq:threshold-volume-difference}, including
for distributions with atoms.

To demonstrate that no universal sign follows, it suffices to
consider the admissible choice $h_0=1$, $h_P=2$, with
\[
 \mathbb P(\varepsilon_t=1)=\mathbb P(\varepsilon_t=-1)=\frac38,
 \quad
 \mathbb P(\varepsilon_t=4)=\mathbb P(\varepsilon_t=-4)=\frac18
\]
For $1<s<2$, enumeration of the sums gives
\[
    A_1(s)=\frac{4-s}{4},\quad
    A_2(s)=\frac{37-11s}{16}
\]
For $4/7<\theta<4/5$, the benchmark spread
$s_0=4\theta/(4-3\theta)$ lies in this interval, and
\[
    \eta_{\mathrm{crit}}
       =1-\frac{4(4-s_0)}{37-11s_0},
    \quad
    \frac{d\eta_{\mathrm{crit}}}{ds_0}
       =-\frac{28}{(37-11s_0)^2}<0
\]
Thus the threshold decreases with toxicity on this interval,
and the screened venue executes more volume at the matching spread.
The Gaussian case proved next gives the opposite signs.
\end{proof}

\begin{proof}[Proof of Proposition \ref{prop:threshold-gaussian}]
Integrating the normal density gives
\[
    A_h(s)=2\int_{s/(\sigma\sqrt h)}^\infty
                     (\sigma\sqrt h\,x-s)\phi(x)\,dx
\]
The identity $\int_a^\infty x\phi(x)\,dx=\phi(a)$ gives
\eqref{eq:gaussian-adverse-selection}.
Normal scaling implies
$A_h(\sqrt{h/h_0}\,s)=\sqrt{h/h_0}\,A_{h_0}(s)$.
Substitution into the break-even equation and uniqueness prove
\eqref{eq:gaussian-spread-scaling}.

Let $f_h$ be the density of $|Z_h|$. For $h_P>h_0$,
\[
    \frac{f_{h_P}(x)}{f_{h_0}(x)}
       =\sqrt{\frac{h_0}{h_P}}\,
          \exp\!\left[
           \frac{x^2}{2\sigma^2}
               \left(\frac1{h_0}-\frac1{h_P}\right)\right],
           \quad x\geq0
\]
It is strictly increasing. Lemma \ref{lem:tail-severity-order}
therefore implies
\[
    \frac{A_{h_P}(s)}{\mathbb P(|Z_{h_P}|>s)}
      >
    \frac{A_{h_0}(s)}{\mathbb P(|Z_{h_0}|>s)}
\]
The derivative and volume identities prove strict increase of
the threshold with $\theta$ and the strict volume comparison.
Common scaling gives independence of $\sigma$.
The low-toxicity limit follows from
$\mathbb E|Z_h|=\sigma\sqrt{2h/\pi}$.

As $\theta\uparrow1$, $s_0\to\infty$. Otherwise, a bounded
subsequence in the break-even equation would imply
$A_{h_0}(s)=0$ at a finite argument, impossible for a Gaussian.
Since $f_{h_0}(x)/f_{h_P}(x)$ is decreasing,
\[
    \frac{A_{h_0}(s)}{A_{h_P}(s)}
       \leq\frac{f_{h_0}(s)}{f_{h_P}(s)}
       \longrightarrow0
       \quad(s\to\infty)
\]
The threshold thus converges to one.
\end{proof}

\begin{proof}[Proof of Proposition \ref{prop:equal-cadence-premium}]
The root is continuous in $\eta$ by Lemma \ref{lem:stop-loss-root}.
Strict decrease of the screened spread proves strict increase
of the premium and its stated endpoints.
Subtract the two break-even equations to obtain
\[
    \Delta s
      =\eta s_0
        -r(\theta,\eta)
          [A_{h_0}(s^*(h_0,\theta,\eta))-A_{h_0}(s_0)]
\]
The integrated-tail identity in Lemma \ref{lem:stop-loss-root}
gives \eqref{eq:equal-cadence-premium}.
For $0<\eta<1$, the integration interval has positive length,
$r(\theta,\eta)>0$, and its integrand is at least
$\mathbb P(|Z_{h_0}|>s_0)$, which is positive because
$A_{h_0}(s_0)=(1-\theta)s_0/\theta>0$. This proves the strict inequalities.

As $\eta\downarrow0$, $\Delta s\downarrow0$. The left limit of
the survival function implies
\[
    \frac1{\Delta s}\int_{s_0-\Delta s}^{s_0}
                     \mathbb P(|Z_{h_0}|>u)\,du
        \longrightarrow\mathbb P(|Z_{h_0}|\geq s_0)
\]
Dividing the premium identity by $\eta$ and taking the limit
gives \eqref{eq:equal-cadence-premium-local}.
\end{proof}

\section{Proofs for the drift extension}
\label{app:drift-proofs}

\begin{proof}[Proof of Proposition \ref{prop:drift-equilibrium}]
For every fixed $\mu$, $|Z_h(\mu)|$ is integrable.
If $\mu\ne0$, its mean is at least $|\mu|h>0$.
If $\mu=0$, conditional Jensen gives
$\mathbb E|M_h|\geq\mathbb E|\xi_t|>0$.
Lemma \ref{lem:stop-loss-root} therefore proves existence,
uniqueness, positivity, and the comparisons in $\theta$ and $\eta$.

To prove joint continuity, observe that
\[
    |A_h(s;\mu)-A_h(\widetilde s;\widetilde\mu)|
       \leq |s-\widetilde s|+h|\mu-\widetilde\mu|
\]
For any convergent parameter sequence with limiting
$\theta\in(0,1)$, roots are locally bounded by
$r(\theta,\eta)(|\mu|h+\mathbb E|M_h|)$.
Every convergent subsequence solves the limiting break-even
equation by the displayed continuity estimate. Uniqueness
identifies its limit, proving convergence of the whole sequence.
This also covers $\eta=1$.
\end{proof}

\begin{proof}[Proof of Proposition \ref{prop:drift-cadence}]
Under symmetry, $M_h$ is symmetric. For fixed $h$ and $s$, set
$\psi_h(a)=\mathbb E[(|M_h+a|-s)_+]$.
It is even and convex, hence nondecreasing for $a\geq0$.
It follows that $A_h(s;\mu)=\psi_h(\mu h)$ is even in $\mu$
and nondecreasing in $|\mu|$.

For the horizon comparison, condition on $M_h$ and apply
Jensen to the next centered increment:
\[
\begin{aligned}
    A_{h+1}(s;\mu)
       &=\mathbb E[\ell_s(M_h+\xi_{t+h}+\mu(h+1))]\\
       &\geq \psi_h(\mu(h+1))
        \geq \psi_h(\mu h)=A_h(s;\mu)
\end{aligned}
\]
The second inequality uses evenness and
$|\mu|(h+1)\geq|\mu|h$.
Loss ordering in Lemma \ref{lem:stop-loss-root} gives the
equilibrium comparisons.

If $\sigma>0$, the additional increment has full support, so
the conditional Jensen inequality is strict, as in the
zero-drift cadence proof. Moreover, $M_h$ has a symmetric,
continuous density $f_{M_h}$ that is positive everywhere.
For $s\geq0$, differentiation under the expectation, followed
by differentiation of its distribution function, gives
\[
    \psi_h''(a)=f_{M_h}(s-a)+f_{M_h}(-s-a)>0
\]
Thus $\psi_h$ is strictly convex and even, so it is strictly
increasing with $|a|$. For $\eta<1$, strict loss ordering
transfers both strict comparisons to the spread.
\end{proof}

\begin{proof}[Proof of Proposition \ref{prop:drift-threshold}]
At the benchmark root,
\[
    A_{h_0}(s_0(\mu);\mu)=\frac{1-\theta}{\theta}s_0(\mu)>0
\]
Proposition \ref{prop:drift-cadence} gives
$A_{h_P}(s_0(\mu);\mu)\geq A_{h_0}(s_0(\mu);\mu)$.
Thus the ratio is well defined, and its threshold lies in $[0,1)$.
At the benchmark spread,
\[
    \pi(h_P,s_0(\mu),\theta,\eta;\mu)
      =\theta A_{h_P}(s_0(\mu);\mu)
             [\eta-\eta_{\mathrm{crit}}(h_P,\theta;\mu)]
\]
Strict increase of profit in $s$ proves all three rankings.
The equal-horizon case and horizon monotonicity follow from
the defining ratio, with strictness when losses strictly increase.

Joint continuity follows from Proposition \ref{prop:drift-equilibrium}
and continuity of the losses, with a positive denominator.
As $\theta\downarrow0$, the root tends to zero because
$s_0(\mu)\leq\theta A_{h_0}(0;\mu)/(1-\theta)$.
Continuity at zero gives \eqref{eq:drift-low-toxicity}.
Finally, common scaling of $(\mu,\xi_t)$ scales $Z_h(\mu)$
and its equilibrium spread by the same factor, and scales both
losses in the ratio equally.
\end{proof}

\begin{proof}[Proof of Proposition \ref{prop:drift-identities}]
For fixed $\mu$, the law of $|Z_h(\mu)|$ does not depend on
$\theta$ or $\eta$. At a non-atom,
$\partial_s A_h(s;\mu)=-S_h(s;\mu)$, and
\[
    \frac{\partial s_0(\mu)}{\partial\theta}
      =\frac{s_0(\mu)}
              {\theta[1-\theta+\theta S_{h_0}(s_0(\mu);\mu)]}
\]
The local root argument in the proof of
Proposition \ref{prop:threshold-toxicity-volume} justifies this
derivative even if there are atoms elsewhere.
Differentiation of the ratio in \eqref{eq:drift-threshold}
gives \eqref{eq:drift-toxicity-derivative}.
Substitution of that same ratio into \eqref{eq:drift-volume}
gives \eqref{eq:drift-volume-comparison}, without differentiating.

For the equal-horizon statement, write $s_h^{\mathrm{ns}}=s^*(h,\theta,0;\mu)$
and $s_h^{\eta}=s^*(h,\theta,\eta;\mu)$. Subtracting their equations gives
\[
    s_h^{\mathrm{ns}}-s_h^{\eta}
       =\eta s_h^{\mathrm{ns}}
          -r(\theta,\eta)[A_h(s_h^{\eta};\mu)-A_h(s_h^{\mathrm{ns}};\mu)]
\]
Lemma \ref{lem:stop-loss-root} proves the integral identity.
Continuity, strict monotonicity, and the endpoint values follow
from Proposition \ref{prop:drift-equilibrium}. For interior
$\eta$, the integral is positive because $s_h^{\eta}<s_h^{\mathrm{ns}}$ and
$S_h(s_h^{\mathrm{ns}};\mu)>0$. Dividing by $\eta$, and using
\[
    \frac1{s_h^{\mathrm{ns}}-s_h^{\eta}}
       \int_{s_h^{\eta}}^{s_h^{\mathrm{ns}}}S_h(u;\mu)\,du
       \longrightarrow\mathbb P(|Z_h(\mu)|\geq s_h^{\mathrm{ns}}),
\]
gives the local expansion, including at atoms.
These arguments do not use symmetry for the equal-horizon claims.

For completeness, symmetric centered innovations still do not
imply a universal sign for the toxicity derivative. The
counterexample in the proof of
Proposition \ref{prop:threshold-toxicity-volume} is already
symmetric at $\mu=0$. Fix a toxicity strictly inside its stated
interval. The equilibrium spread lies a positive distance from
every relevant atom. Under a sufficiently small nonzero drift,
continuity keeps the spread away from those atoms, the survival
probabilities remain unchanged, and the loss functions and the
derivative formula vary continuously. The negative derivative
therefore persists for nonzero drifts with symmetric centered
innovations. The Gaussian case below supplies the opposite sign.
\end{proof}

\begin{proof}[Proof of Proposition \ref{prop:drift-gaussian}]
Use the identity
$(|z|-s)_+=(z-s)_++(-z-s)_+$.
For a normal variable of mean $m$ and standard deviation $v$,
direct integration gives
\[
    \mathbb E[(X-s)_+]
      =v\phi((s-m)/v)+(m-s)\overline\Phi((s-m)/v)
\]
Apply it to $Z_h(\mu)$ and $-Z_h(\mu)$ to obtain
\eqref{eq:drift-gaussian-loss}.

The density of the absolute innovation, for $x\geq0$, is
\[
    f_h(x;\mu)
      =\frac{\sqrt{2/\pi}}{\sigma\sqrt h}
          \exp\!\left[-\frac{x^2}{2\sigma^2h}
                       -\frac{\mu^2h}{2\sigma^2}\right]
          \cosh\!\left(\frac{\mu x}{\sigma^2}\right)
\]
The hyperbolic-cosine term is the same at both horizons. Hence
\begin{equation}
    \frac{f_{h_P}(x;\mu)}{f_{h_0}(x;\mu)}
      =\sqrt{\frac{h_0}{h_P}}\,
        \exp\!\left[
         \frac{x^2}{2\sigma^2}
              \left(\frac1{h_0}-\frac1{h_P}\right)
           -\frac{\mu^2(h_P-h_0)}{2\sigma^2}\right]
    \label{eq:drift-density-ratio}
\end{equation}
For $h_P>h_0$, this ratio is strictly increasing in $x$,
regardless of the sign or magnitude of $\mu$.
Lemma \ref{lem:tail-severity-order} proves
\[
    \frac{A_{h_P}(s;\mu)}{S_{h_P}(s;\mu)}
       >
    \frac{A_{h_0}(s;\mu)}{S_{h_0}(s;\mu)}
\]
The identities in Proposition \ref{prop:drift-identities}
give a strict increase of the threshold with $\theta$ and the
strict volume comparison.

For fixed $\mu$, Gaussian tails imply $A_{h_0}(s;\mu)>0$ at
every finite $s$. The benchmark equation therefore forces
$s_0(\mu)\to\infty$ as $\theta\uparrow1$.
The reciprocal of \eqref{eq:drift-density-ratio} is decreasing
and tends to zero. Consequently
\[
    0<\frac{A_{h_0}(s;\mu)}{A_{h_P}(s;\mu)}
      \leq\frac{f_{h_0}(s;\mu)}{f_{h_P}(s;\mu)}
      \longrightarrow0
\]
This proves the high-toxicity limit.

Setting $s=0$ in \eqref{eq:drift-gaussian-loss} gives
$A_h(0;\mu)=\kappa_h(\mu)$, proving the low-toxicity formula.
To establish its monotonicity, define
\[
    g(a)=2\phi(a)+a[2\Phi(a)-1],\quad a\geq0
\]
Then $\kappa_h(\mu)=\sigma\sqrt h\,g(|\mu|\sqrt h/\sigma)$,
$g'(a)=2\Phi(a)-1$, and $g''(a)=2\phi(a)$.
The elasticity $e(a)=a g'(a)/g(a)$ satisfies, for $a>0$,
\[
    e'(a)
      =\frac{2\phi(a)[\,2\Phi(a)-1+a g(a)\,]}{g(a)^2}>0
\]
For $b=|\mu|/\sigma>0$,
\[
    \frac{d}{d\log b}
       \log\frac{\kappa_{h_P}(\mu)}{\kappa_{h_0}(\mu)}
       =e(b\sqrt{h_P})-e(b\sqrt{h_0})>0
\]
Continuity covers $b=0$. Since $g(0)=\sqrt{2/\pi}$ and
$g(a)/a\to1$, the two stated endpoint limits follow.

Finally, with $N$ standard normal,
\[
    \frac{Z_h(\mu)}{\sigma\sqrt{h_0}}
       \overset d=
       \frac{\mu\sqrt{h_0}}{\sigma}\frac{h}{h_0}
          +\sqrt{\frac{h}{h_0}}\,N
\]
Homogeneity of the loss and uniqueness of the root establish
the parameter reduction for the threshold.
For fixed nonzero $\mu$, normalizing the spread by $\sigma\sqrt h$
leaves a break-even equation whose absolute normal location is
$|\mu|\sqrt h/\sigma$. Its root strictly increases with that
location when $\eta<1$, by Proposition \ref{prop:drift-cadence}
applied to a fixed unit-variance normal law.
Thus square-root scaling fails at nonzero drift, except for
the identically zero spread under perfect screening.
\end{proof}

\begin{proof}[Proof of Proposition \ref{prop:drift-local}]
Gaussian loss is smooth in $(s,\mu)$ near $(s^{\mathrm{nd}},0)$.
Put $v=\sigma\sqrt h$. Differentiating
\eqref{eq:drift-gaussian-loss} gives
\[
    A_{h,\mu}(s;\mu)
       =h\left[
           \overline\Phi\!\left(\frac{s-\mu h}{v}\right)
          -\overline\Phi\!\left(\frac{s+\mu h}{v}\right)
         \right],
\]
so
\[
    A_{h,\mu}(s;0)=0,\quad
    A_{h,\mu\mu}(s;0)=\frac{2h^2}{v}\phi(s/v).
\]
The spread equation has a derivative in $s$ equal to
$1+rS_h(s;\mu)>0$, so the implicit function theorem gives
a smooth equilibrium root. Its first derivative at zero
vanishes, and twice differentiating yields.
\[
    \left.\frac{\partial^2s^*}{\partial\mu^2}\right|_{\mu=0}
       =\frac{2r h^2\phi(s^{\mathrm{nd}}/v)}
                    {v[1+rS_h(s^{\mathrm{nd}};0)]}
\]
Taylor expansion, with the factor $1/2$ on this derivative,
gives \eqref{eq:drift-local-expansion}.
\end{proof}

\begin{proof}[Proof of Proposition \ref{prop:drift-dominated}]
The positive-part absolute-value payoff is $1$-Lipschitz in
its price argument. Uniformly in $s\geq0$,
\[
    \left|
      \mathbb E[(|\mu h+cM_h|-s)_+]-(|\mu|h-s)_+
    \right|
       \leq c\,\mathbb E|M_h|\longrightarrow0
\]
The limiting break-even function is
$s-r(|\mu|h-s)_+$, which is strictly increasing and has the
unique root $r|\mu|h/(1+r)$. Uniform convergence of the loss
functions implies convergence of their roots: for example,
the break-even functions increase at least as fast as $s$,
so the distance between a root and the limiting root is at
most the absolute function error at the limiting root.
Substituting $r=\theta(1-\eta)/(1-\theta)$ gives the stated spread.

At the unscreened benchmark,
$s_{0,c}\to\theta|\mu|h_0$. Uniform loss convergence and the
Lipschitz property in $s$ imply
\[
    A_{h_0,c}(s_{0,c};\mu)\to|\mu|h_0(1-\theta),
    \quad
    A_{h_P,c}(s_{0,c};\mu)\to|\mu|(h_P-\theta h_0)>0
\]
Substituting into the threshold ratio gives
$1-h_0(1-\theta)/(h_P-\theta h_0)
 =(h_P-h_0)/(h_P-\theta h_0)$.
\end{proof}

\begin{proof}[Proof of Proposition \ref{prop:drift-recentering}]
For a log center $p_t+\gamma_h$, the linearized profit from an order
is $s-D_i(Z_h(\mu)-\gamma_h)$. An informed order executes precisely
when $|Z_h(\mu)-\gamma_h|>s$, with direction given by the sign of
that pricing error. At $\gamma_h=\mu h$, the error is $M_h$.
Since its mean is zero and uninformed direction is independent
of it, uninformed expected profit is $s$ even without balanced
directions. The loss, participation, and break-even equations
are exactly those of the zero-drift model for $\xi_t$.

For the optimizing assertion, balanced uninformed flow makes
its expected contribution $(1-\theta)s$ independent of $\gamma_h$.
The informed loss is
\[
    \mathbb E[(|M_h+\mu h-\gamma_h|-s)_+]
\]
Under symmetry, this is an even convex function of $\mu h-\gamma_h$,
so it is minimized at zero for every $s$. Loss ordering implies
the smallest break-even spread is attained there.
With a Gaussian component, the strict convexity established in
the proof of Proposition \ref{prop:drift-cadence} makes this
minimizer unique. For $\eta<1$, the informed-loss coefficient
is positive, so any other center has strictly higher loss at
the minimizing equilibrium spread and a strictly larger root.
\end{proof}

\section{Why drift alone does not preserve the cadence ordering}
\label{app:drift-counterexample}

Take i.i.d.\ increments with
$\mathbb P(\varepsilon_t=1)=9/10$ and
$\mathbb P(\varepsilon_t=-4)=1/10$. Their drift is $\mu=1/2$,
and the centered increments are not symmetric. At $s=3$,
\[
    A_1(3;\mu)=\frac1{10},
    \quad
    A_2(3;\mu)=\frac1{20}
\]
With $\theta=30/31$ and $\eta=0$, $r=30$ and the first-horizon
spread is $3$. On $2<s<3$, the second-horizon loss is
$A_2(s;\mu)=(62-19s)/100$, giving
$s^*(2,\theta,0;\mu)=186/67<3$.
This is a counterexample within the balanced-flow model.
Multiplying all increments and spreads by any positive constant
preserves it, so it can be placed within the small-return regime.

\section{Screening with false positives and false negatives}
\label{app:screening-errors}

Retain the zero-drift model of Section \ref{sec:model}, including its
linearized-profit approximation, potential informed share $\theta\in(0,1)$,
and stale-loss function $A_h(s)=\mathbb E[(|Z_h|-s)_+]$. Let $\alpha\in[0,1)$ be the
false-positive rate, the probability of rejecting an uninformed order, and let
$\beta\in[0,1]$ be the false-negative rate, the probability of admitting an
informed order. Thus, uninformed and informed survival probabilities are
$1-\alpha$ and $\beta$, respectively. In the notation of the baseline,
\[
 \mathbb P(B_i=1\mid U,D_i,Z_h)=1-\alpha,
 \quad
 \mathbb P(B_i=1\mid I,Z_h)=\beta
\]
The baseline preserves all uninformed flow, making the screening rate a direct measure of the losses removed. With classification errors, rejecting an order can remove either informed losses or uninformed revenue. The appropriate measure of protection is therefore the survival of informed opportunities relative to that of uninformed opportunities. This extension retains random selection within each type, so the payoff distribution of admitted orders is unchanged conditional on type.

\begin{proposition}[Relative survival and the break-even spread]
\label{prop:relative-survival}
Assume $\mathbb E|Z_h|<\infty$. The unique nonnegative break-even half-spread
under the two-error screen solves
\begin{equation}\label{eq:two-error-break-even}
 (1-\theta)(1-\alpha)s=\theta\beta A_h(s)
\end{equation}
Define
\begin{equation}\label{eq:effective-screening}
 \eta_{\mathrm{eff}}:=1-\frac{\beta}{1-\alpha}
\end{equation}
If $\beta\leq 1-\alpha$, the spread is exactly
$s^*(h,\theta,\eta_{\mathrm{eff}})$ from the baseline model. Hence the two
error rates affect the spread only through their relative survival rate.
\end{proposition}

\begin{proof}
Per unit of potential flow, admitted uninformed flow earns expected margin
$(1-\theta)(1-\alpha)s$, while admitted informed flow generates expected
loss $\theta\beta A_h(s)$. Dividing their zero-profit condition by
$1-\alpha>0$ gives the baseline equation with
$1-\eta$ replaced by $\beta/(1-\alpha)$. Its left side is strictly increasing
in $s$, its right side is continuous and nonincreasing, and the difference
is nonpositive at zero and tends to infinity. The root is therefore unique.
\end{proof}

The restriction $\beta\leq1-\alpha$ says that screening admits informed
flow no more readily than uninformed flow; it is precisely what ensures
$\eta_{\mathrm{eff}}\in[0,1]$. If $\beta>1-\alpha$, the same algebra holds
but $\eta_{\mathrm{eff}}<0$: the screen worsens selection and lies outside
the baseline screening range. Thus the one-parameter formulation is a
normalization for the spread result under the maintained screening
assumption, not a claim that every classifier is summarized by its
unconditional error rates.

In particular, retain $s_0=s^*(h_0,\theta,0)$ and $h_P\geq h_0$ from
Theorem \ref{prop:screening-threshold}. Its proof ensures
$A_{h_P}(s_0)\geq A_{h_0}(s_0)>0$. The screened mechanism at $h_P$
has a break-even spread no larger than
$s_0$ exactly when
\begin{equation}\label{eq:two-error-threshold}
 \frac{\beta}{1-\alpha}\leq\frac{A_{h_0}(s_0)}{A_{h_P}(s_0)},
 \quad\text{equivalently}\quad
 \eta_{\mathrm{eff}}\geq\eta_{\mathrm{crit}}(h_P,\theta)
\end{equation}
Equality gives equal spreads. This follows by evaluating the increasing
profit function in \eqref{eq:two-error-break-even} at $s_0$.

\runhead{Volume} Executed volume per unit of potential flow is
\begin{equation}\label{eq:two-error-volume}
\begin{aligned}
 &(1-\theta)(1-\alpha)+\theta\beta\,\mathbb P(|Z_h|>s)\\
 &\quad=(1-\alpha)\big[(1-\theta)+\theta(1-\eta_{\mathrm{eff}})
                   \mathbb P(|Z_h|>s)\big]
\end{aligned}
\end{equation}
For $\eta_{\mathrm{eff}}\in[0,1]$, this is
$(1-\alpha)V(h,s,\theta,\eta_{\mathrm{eff}})$ in the baseline notation.
Expected interval volume is $h$ times this expression. Screens with the same $\eta_{\mathrm{eff}}$ therefore
support the same spread but can generate different executed volumes.

\section{Heterogeneous quote ages}
\label{app:execution-ages}

Use the continuous-time zero-drift specification in Section \ref{sec:model},
so $Z_a=p_{t+a}-p_t$ is defined at every age $a\geq0$. In discrete time,
restrict ages to integers; fractional-age losses require an additional
specification. Keep the quote fixed over $[0,h]$ and reset it at $h$.
Let $T_h\in[0,h]$ be the age at a potential informed execution opportunity,
with law $G_h$ independent of $s$ and $\eta$. Age is independent of the price
process and sampled before the execution decision, not conditional on a
completed trade. At age $a$, an informed trader observes $Z_a$ and executes
precisely when $|Z_a|>s$.

Screening remains independent of age and innovations conditional on type,
and preserves uninformed directions. Uninformed opportunity ages are also
independent of the price process; conditional on age, direction is
independent of the innovation. Thus $\mathbb E[Z_a]=0$ gives uninformed
expected margin $s$ without requiring balanced directions. Retain the
potential-flow normalization and linearized profits.

Define the execution-age-averaged stale loss by
\begin{equation}\label{eq:age-averaged-loss}
 \bar A_h(s):=\mathbb E[(|Z_{T_h}|-s)_+]
       =\int_{[0,h]} A_a(s)\,dG_h(a),
 \quad \mathbb E|Z_{T_h}|<\infty
\end{equation}
The endpoint baseline is recovered when $T_h=h$ almost surely.

\begin{proposition}[The threshold with heterogeneous quote ages]
\label{prop:age-threshold}
For $\theta\in(0,1)$ and $\eta\in[0,1]$, the unique nonnegative break-even
half-spread $\bar s^*(h,\theta,\eta)$ solves
\begin{equation}\label{eq:age-break-even}
 (1-\theta)\bar s^*
       =\theta(1-\eta)\bar A_h(\bar s^*)
\end{equation}
For $h_P\geq h_0$, define the benchmark within this extension as
$\bar s_0:=\bar s^*(h_0,\theta,0)$.
If $\bar A_{h_P}(\bar s_0)>0$, then
\begin{equation}\label{eq:age-threshold}
 \bar s^*(h_P,\theta,\eta)\leq\bar s_0
 \quad\Longleftrightarrow\quad
 \eta\geq\bar\eta_{\mathrm{crit}}(h_P,\theta)
 :=1-\frac{\bar A_{h_0}(\bar s_0)}{\bar A_{h_P}(\bar s_0)}
\end{equation}
Equality of spreads holds exactly at equality in the threshold condition.
If $\bar A_{h_P}(\bar s_0)\geq\bar A_{h_0}(\bar s_0)>0$, then
$\bar\eta_{\mathrm{crit}}\in[0,1)$; the threshold is strictly positive when
the loss inequality is strict.
\end{proposition}

\begin{proof}
The positive-part payoff is continuous and $1$-Lipschitz in $s$, so
$\bar A_h$ is continuous and nonincreasing. Consequently
$\bar\pi_h(s)=(1-\theta)s-\theta(1-\eta)\bar A_h(s)$ is strictly increasing,
$\bar\pi_h(0)\leq0$, and $\bar\pi_h(s)\to\infty$. This proves existence and uniqueness.
At the fast benchmark's root,
\[
 \bar\pi_{h_P}(\bar s_0)=\theta\big[
 \bar A_{h_0}(\bar s_0)-(1-\eta)\bar A_{h_P}(\bar s_0)\big]
\]
The slow mechanism's root is no larger than $\bar s_0$ precisely when
this expression is nonnegative, yielding \eqref{eq:age-threshold}.
\end{proof}

The benchmark $\bar s_0$ must be recomputed under the age distribution.
The root is positive if $\eta<1$ and $\bar A_h(0)>0$.
If the denominator in \eqref{eq:age-threshold} is zero, the profit
comparison directly implies a spread no larger than the benchmark for
every feasible $\eta$.

\runhead{Ordering stale exposure} The condition $h_P>h_0$ alone does not order averaged losses. A sufficient
condition is that $a\mapsto A_a(s)$ is nondecreasing for every $s\geq0$
and that $T_{h_P}$ first-order stochastically dominates $T_{h_0}$. Then
$\bar A_{h_P}(s)\geq\bar A_{h_0}(s)$ for all $s$. The first condition
holds for the maintained integrable martingale log price by conditional
Jensen, since $z\mapsto(|z|-s)_+$ is convex. These are conditions on stale
exposure; no order-arrival or routing model is required.

\runhead{Uniform ages} For exogenous $T_h\sim\mathrm{Unif}[0,h]$,
\begin{equation}\label{eq:uniform-age-loss}
 \bar A_h(s)=\frac1h\int_0^h A_a(s)\,da
             =\int_0^1 A_{hu}(s)\,du
\end{equation}
A common uniform draw orders ages across horizons, so the martingale
property gives nondecreasing averaged losses in $h$ and
$\bar A_h(s)\leq A_h(s)$. In the Gaussian case of
Proposition \ref{prop:threshold-gaussian}, integration of
\eqref{eq:gaussian-adverse-selection} at zero spread gives
\[
 \bar A_h(0)=\frac23\sigma\sqrt{\frac{2h}{\pi}}
            =\frac23 A_h(0)
\]
This factor applies at $s=0$; it is not a constant adjustment to the
break-even spread or the screening threshold.

Both extensions can be applied together: replace $A_h(s)$ by
$\bar A_h(s)$ in \eqref{eq:two-error-break-even} and use
$\eta_{\mathrm{eff}}$ in \eqref{eq:age-threshold}. The resulting comparison
continues to hold at the same exogenous potential-flow composition.

\section{Fee income and fast adverse selection}
\label{app:extended}

One further analysis is reported for completeness; it is not load-bearing for the paper's claims. It concerns liquidity-provider economics (Figure \ref{fig:lp}). Measured as the volume-weighted five-second markout and set against the posted fee, post-trade adverse selection absorbs 0.2\% to 5.3\% of the fee tier, and every pool is net-positive on this measure. Because the stale-quote loss is paid at execution, the figure also reports the volume-weighted half effective spread against the benchmark mid, the revenue a provider keeps at the moment of the fill: 45\% of the posted fee in the v3 1\,bp pool, 14\% in the v2 pool, and 1\% to 3\% in the v3 5\,bps and 30\,bps pools. This bounds only the fast, arbitrage-driven component of provider losses; rebalancing losses beyond five seconds sit outside the measurement, so this does not speak to loss-versus-rebalancing.

\begin{center}
\begin{minipage}{\textwidth}
\centering
\includegraphics[width=0.8\textwidth]{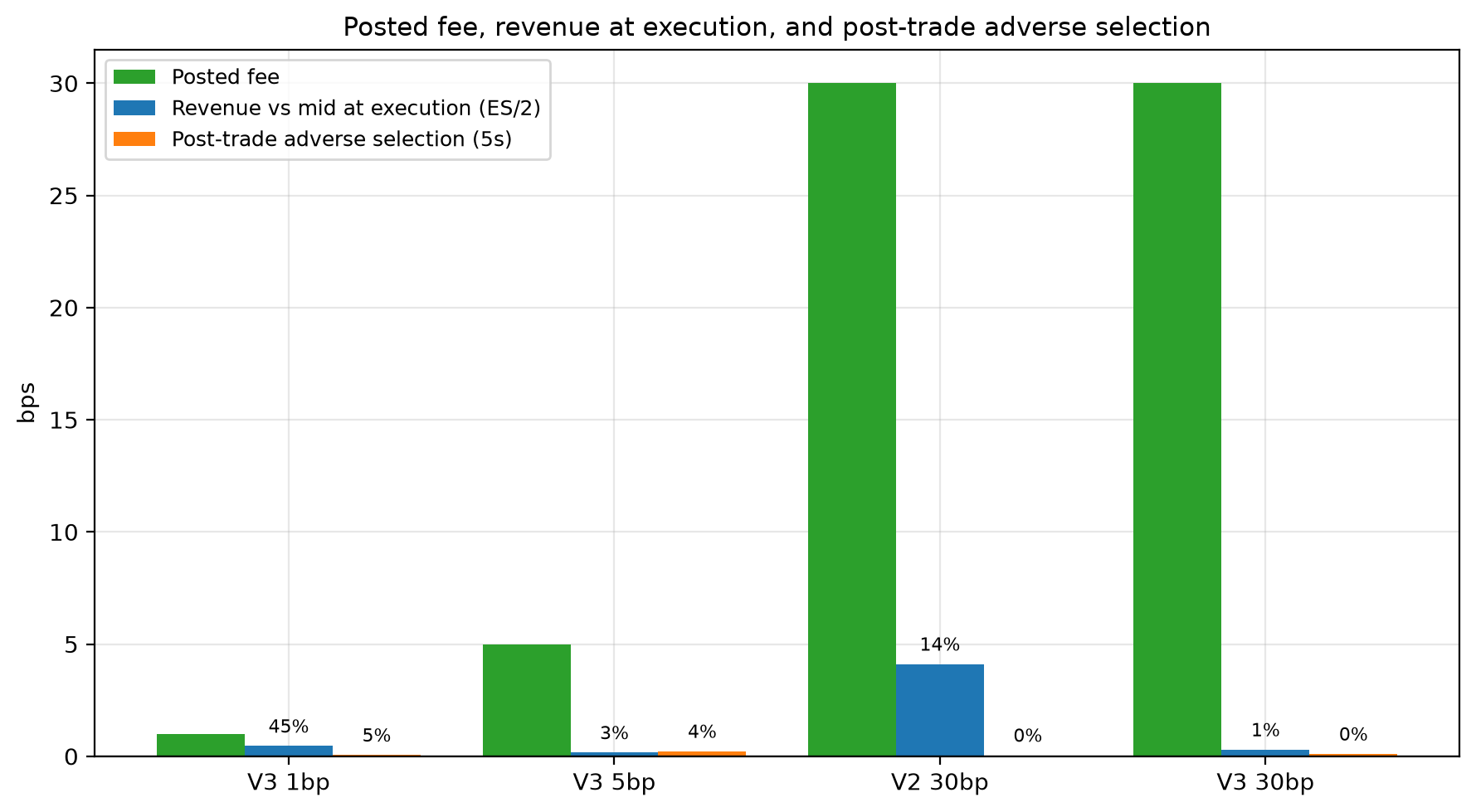}
\makeatletter\def\@captype{figure}\makeatother
\caption{Posted fee, revenue at execution, and five-second adverse selection across the Uniswap pools. Post-trade adverse selection absorbs 0.2-5.3\% of the posted fee, and every pool is net-positive on this measure; the volume-weighted half effective spread against the benchmark mid, the revenue kept at execution, is 1-45\% of the fee. The comparison bounds the fast arbitrage-driven component only; rebalancing losses beyond five seconds are not captured.}
\label{fig:lp}
\end{minipage}
\end{center}

\end{document}